\documentstyle{mn}
\input{epsf}

\title[Characteristics of simulated walls]
{Statistical characteristics of simulated walls}

\author[Demia\'nski et al.]
       {M. Demia\'nski$^{1,2}$,  A.G. Doroshkevich$^{3,4}$,
        V.M\"uller$^5$ \& V.Turchaninov$^4$\\
        $^1$ Institute of Theoretical Physics,
                       University of Warsaw,
                       00-681 Warsaw, Poland\\
        $^2$ Department of Astronomy, Williams College,
                       Williamstown, MA 01267, USA\\
        $^3$ Theoretical Astrophysics Center,
                       Juliane Maries Vej 30,
                       DK-2100 Copenhagen \O, Denmark\\
        $^4$ Keldysh Institute of Applied Mathematics,
                       Russian Academy of Sciences,
                       125047 Moscow,  Russia\\
        $^5$ Astrophysikalisches Institute Potsdam,
	             An der Sternwarte 16, Potsdam, D-14482 Germany}

\date{Accepted ...,
      Received 2000 March ...;
	in original form 1999 ...}
\begin{document}
\maketitle

\begin{abstract}
The large scale matter distribution in three different simulations
of CDM models is investigated and compared with  corresponding
results of the Zel'dovich theory of nonlinear gravitational
instability. We show that the basic characteristics of wall-like
structure elements are well described by this theory, and that
they can be expressed by the cosmological parameters and a few
spectral moments of the perturbation spectrum. Therefore the
characteristics of such elements provide reasonable estimates
of these parameters. We show that the compressed matter is
relaxed and gravitationally confined, what manifests itself
in the existence of walls as (quasi)stationary structure elements
 with life time  restricted by their disruption into high density
 clouds.

The matter distribution is investigated both in the real and
redshift spaces. In both cases almost the same particles form the
walls, and we estimate differences in corresponding  wall
characteristics. The same methods are applied to several mock
catalogues of 'galaxies' what allows us to characterize a large
scale bias between the spatial distribution of dark matter and
of simulated `galaxies'.
\end{abstract}


\section{Introduction}

Over the past decade immense progress was achieved in the
investigation of the large scale matter distribution. Now the
galaxy distribution is studied up to the redshift $z\sim 3$
(Steidel at al. 1996). At smaller redshifts the
analysis of rich galaxy surveys with an effective depth $\sim$
(200 -- 400)$h^{-1}$Mpc, such as the Durham/UKST Galaxy Redshift
Survey (Ratcliffe et al. 1996), and the Las Campanas Redshift
Survey (Shectman et al. 1996), have established the existence
of wall-like structure elements as a typical phenomenon in
the visible galaxy distribution incorporating $\sim$ (40 -- 50)\%
of galaxies (Doroshkevich et al. 1996, hereafter LCRS1;
Doroshkevich et al. 1999a, hereafter LCRS2; Doroshkevich et al.
2000, hereafter DURS). The wall-like structure elements with a
typical diameter $\sim$ (30 -- 50)$h^{-1}$Mpc surround low-density
regions with a similar typical diameter $\sim$ (50 -- 70)$h^{-1}$
Mpc. Within the wall-like structures, the observed galaxy
distribution is also inhomogeneous (see, e.g., Fig. 5 of Ramella
et al. 1992), and galaxies are concentrated in high density
clumps and filaments.

The galaxies occupying low density regions are concentrated within
a random network of filaments. Filaments incorporate $\sim$ 50\%
of galaxies and are clearly seen in many redshift surveys (see,
e.g., de Lapparent, Geller \& Huchra 1988).
These results extend the range of investigated scales in the galaxy
distribution up to $\sim$ 100$h^{-1}$Mpc. Further progress in the
study of the observed large scale galaxy distribution could be
reached with the 2dF redshift survey (Colless 1998; Cannon 1998)
and the Sloan Digital Sky Survey (Loveday \& Pier 1998; Maddox 1998).

The formation and evolution of structure on large scales are
investigated in numerous simulations (see, e.g., Cole et al.,
1997, 1998; Jenkins et al. 1998; Governato et al. 1998;
M\"uller et al. 1998; Doroshkevich et al. 1999b, hereafter DMRT).
These simulations
are performed in large boxes ($\sim$ 350 -- 500$h^{-1}$Mpc)
and reproduce the main properties of the  observed large
scale matter distribution. In particular, they confirm
formation of large wall-like matter condensations
due to a nonlinear anisotropic matter compression on a typical
scale $\sim$ (20 -- 30)$h^{-1}$Mpc that is about one half
of the typical wall separation.

The statistical characteristics of wall formation are described by an
approximate theoretical model (Lee \& Shandarin 1998; Demia\'nski ~\&~
Doroshkevich 1999a, b, hereafter DD99) based on the Zel'dovich
nonlinear theory of gravitational instability (Zel'dovich 1970, 1978;
Shandarin ~\&~ Zel'dovich 1989). This approach relates the structure
parameters with the main parameters of the underlying cosmological
scenario and the initial power spectrum of density perturbations. The
impact of large scale perturbations is found to be important
throughout all evolutionary stages and some statistical
characteristics of structure elements -- filaments and walls formed
in the course of nonlinear evolution -- are directly connected with
the parameters of these perturbations.  Another theoretical model of
large scale structure formation was  discussed in Bond, Kofman
\& Pogosyan (1996).

The simulated large scale matter distribution does not exactly
reproduce the theoretical expectations due to the influence of some
essential factors, the most important ones are the small scale
clustering and relaxation of compressed matter, and the large scale
matter flow within sheet-like structure elements. Thus, compression
of matter along one of the transversal directions transforms sheet-like
elements into filaments, while expansion of matter in both transversal
directions results in the erosion of pancakes. The disruption of walls
and the small scale clustering of compressed matter substantially
accelerate the relaxation and are responsible for strong matter
concentration within walls. This is apparent from the isotropy of
velocity dispersion within walls noticed in DMRT.

The combined influence of these (and other) factors complicates the
statistical description of the large scale matter distribution at late
evolutionary stages, what is typical for the final evolutionary stages
of the standard COBE-normalized CDM (SCDM) model with $\Omega_m=1$.
For low density models, such as the open CDM (OCDM) model and the
$\Lambda$CDM model with $\Omega_{\Lambda}>\Omega_{m}$, the situation is
not so complex, and some statistical characteristics of structure can
be successfully compared with the approximate theoretical expectations.

The investigation of wall-like massive structure elements is more
promising in this respect because walls represent the first step
in the process of structure formation and, so, hold more
information about characteristics of the initial matter flow. Such
walls are observed as superclusters of galaxies similar to the Great
Wall (de Lapparent et al. 1988) and the Pisces-Perseus supercluster
(Giovanelli \& Haynes 1993). In simulations such wall-like structure
elements are also easily identified because of their relatively high
overdensity.  Samples of such elements were
investigated in DMRT and LCRS2. The connection
between properties of walls and the amplitude and the spectrum of
initial perturbations was discussed in DD99, and some of these
results can be compared with measured properties of simulated
wall-like structure elements. Examples considered in DD99 and
DMRT had rather illustrative character, but they seem to be
quite promising.

Here we will compare more accurately some of the expected and measured
characteristics of wall-like matter condensations. We concentrate our
attention on the physical aspects of the formation and evolution of
the large scale matter distribution in order to better understand
these processes and the phenomenon of wall-like matter
condensations. Both theoretical and numerical estimates are inevitably
approximate, but nevertheless, such comparison allows us to test the
theoretical conclusions, to reveal and illustrate the influence of
essential factors mentioned above, and to examine the abilities of
statistical methods used to describe the large scale matter
distribution.

These methods allow us to reveal, in particular, some differences in
characteristics of the large scale matter distribution in the real and
redshift spaces.  Various aspects of this problem were widely discussed
during the past decade (see, e.g., Kaiser 1987; McGill 1990 a; Davis,
Miller \& White 1997; Hamilton 1998; Melott et al.  1998; Hui, Kofman
\& Shandarin 1999, Tadros et al.  1999).  Here we show that the
differences between characteristics of walls in the real and redshift
spaces depend on the basic cosmological model and increase during the
cosmic evolution.  Characteristics of walls in the real and redshift
spaces are almost identical for the low-density models, but they differ
more strongly for the SCDM model.

We do not discuss the application of these methods to the
observed galaxy catalogues, what is a much harder problem, due
to the strong influence of selection effects and other
factors. We will consider this problem in the future.

This paper is organized as follows: In Sec. 2 the basic notations are
introduced. In Sec. 3 the statistical characteristics of wall-like
structure elements in the Zel'dovich theory are presented. In Sec. 4
we consider the methods used to measure the required characteristics
of matter distribution. Our results are presented in Secs. 5 ~\&~6
where they are also compared with the theoretical expectations. Sec. 7
contains summary and a short discussion of our main results.  Some
technical details are given in Appendixes A.

\section{Statistical characteristics of large scale structure}

It is generally recognized that the formation of observed large scale
structure is driven by the middle part of the power spectrum, $p(k)$,
with $0.2h{\rm Mpc}^{-1}\geq k\geq 0.01h{\rm Mpc}^{-1}$ ($k$ is the
comoving wave number), and it is weakly sensitive to the small and
large scale perturbations.  In many publications authors use an
artificial smoothing of the spectrum to describe this process (see,
e.g., Bardeen et al.  1986, hereafter BBKS, Coles et al.  1993).
However, as was shown in DD99 it is possible to avoid this artificial
smoothing if the process of structure formation is described in terms
of the displacement, $S_i({\bf q})$, and velocity rather than density
field.

Indeed, in contrast with the density field, the statistical
characteristics of displacements are weakly sensitive to the small and
large scale perturbations and are reasonably well described by the
middle part of the initial power spectrum. Even the strong nonlinear
matter clustering does not significantly influence the main
characteristics of displacements and, so, such (approximate)
description of structure holds during long period of structure
evolution. Of course, this approach cannot describe the formation
of gravitationally confined walls and their disruption into a system
of high density clouds.

Bearing in mind these comments we will describe the structure
parameters using characteristics directly connected with the
displacement. One of them is the large scale amplitude of
perturbations measured by the dispersion of displacements,
$$\sigma_s^2(z) = {1\over 2\pi^2}\int_0^\infty p(z,k)dk,
						\eqno(2.1)$$
Other convenient parameter is the coherent length of the displacement 
and velocity fields, $l_v$, expressed through the moment $m_{-2}$ 
of the initial power spectrum, $p(k)$. Suitably defined coherent  
length $l_v$ provides simple expressions for the correlation functions 
of these fields and the basic characteristics of the large 
scale structure (DD99 and Sec. 3).

\subsection{The Zel'dovich approximation}

The Zel'dovich theory connects the Eulerian, $r_i$, and the
Lagrangian, $q_i$, coordinates of fluid elements (particles)
by the expression
$$r_i = (1+z)^{-1}[q_i - B(z)S_i({\bf q})],\eqno(2.2)$$
$$S_i({\bf q})=\partial\Phi({\bf q})/\partial q_{i},$$
where $z$ denotes the redshift, $B(z)$ describes growth of
perturbations in the linear theory, and the random vector
$S_i$ or the random potential $\Phi$ characterize the spatial
distribution of perturbations. The Lagrangian coordinates of
a particle, $q_i$, are its unperturbed comoving coordinates.

The velocity of a particle can be found from (2.2) as
$$u_i({\bf q},z) ={dr_i\over dt} = {H(z)\over 1+z}[q_i-
(1+\beta)B(z)S_i({\bf q})],$$
$$\beta(z)= -{1+z\over B}{dB(z)\over d z},\eqno(2.3)$$
$$H(z) = H_0\sqrt{\Omega_m(1+z)^3+(1-\Omega_m-\Omega_\Lambda)
(1+z)^2+\Omega_\Lambda},$$
where H is the Hubble constant ($H_0=100\cdot h$ km/s/Mpc).
Analytical fits for the functions $B(z)$ and $\beta(z)$ were given
in DD99. Approximately, at $z\ll 1$, we have
$$B(0)=1,\quad \beta(0)\approx {2.3\Omega_m\over 1+1.3\Omega_m}.
			\eqno(2.4)$$

\subsection{Main structure characteristics for the CDM-like
power spectrum}

The standard CDM-like power spectrum with a Harrison -- Zel'dovich
large scale asymptote
$$p_{cdm}(k) = A(z) kT^2(k/ k_0),\quad k_0 = \Gamma\,h\,
{\rm Mpc}^{-1},\eqno(2.5)$$
$$\Gamma = \sqrt{1.7 \rho_\gamma\over\rho_{rel}}\,\Omega_mh,$$
can be taken as a reasonable approximation of the initial power
spectrum used in Zel'dovich' theory. Here $A(z)$ is the amplitude
of perturbations, $T(x)$ is a transfer function and
$\rho_\gamma~\&~\rho_{rel}$ are the densities of CMB photons and
relativistic particles (photons, neutrinos etc.). For this
spectrum the parameters $l_v$ and $\sigma_s$ are expressed through
the spectral moments, $m_j$, as follows:
$$l_v^{-2} = \int_0^\infty kT^2(k/k_0)dk = m_{-2}k_0^2,
						\eqno(2.6)$$
$$\sigma_s^2 \equiv {1\over 2\pi^2}\int_0^\infty p_{cdm}(k)dk=
{A(z)\over 2\pi^2}k_0^2m_{-2} = {A(z)\over 2\pi^2 l_v^2}$$
$$m_j=\int_0^\infty x^{3+j}T^2(x)dx,\quad
m_{-2}=\int_0^\infty xT^2(x)dx$$
For the CDM transfer function (BBKS) $m_{-2}=0.023$, and the
expressions for the scale $l_v$ and the characteristic masses
of DM and baryonic components associated with the scale $l_v$
can be written more explicitly as
$$l_v\approx {6.6\over\Gamma}~\sqrt{0.023\over m_{-2}}~
h^{-1}{\rm Mpc},		                     \eqno(2.7)$$
$$M_v = {4\pi\over 3}<\rho> l_v^3\approx {2\cdot 10^{14}
M_\odot\over \Gamma^2 h^2},\quad M_b^{(0)} = {\Omega_b\over
\Omega_m}M_v.$$
Here $\Omega_b$ is the dimensionless mean density of the baryonic
component. The same characteristic scale, $l_v$, as given by
(2.7) can be used for the structure description as long as the
Zel'dovich theory can be applied.

More details can be found in DD99. The same approach
can be used for other power spectra as well.

\subsection{The amplitude of large scale perturbations}

The large scale amplitude of perturbation as measured by $A(z)$ 
in (2.5) and $\sigma_s$ (2.1) can be successfully used to describe 
the structure evolution in the framework of the Zel'dovich theory.
As was shown in DD99 it is convenient to use -- together with
$\sigma_s$ -- an effective dimensionless `time', $\tau(z,
\Omega_m,h)$,
$$\tau(z) = {\sigma_s\over\sqrt{3}l_v}
					\eqno(2.8)$$
which is proportional to the large scale amplitude of
perturbations and describes suitably the evolutionary stage
reached in the model. This 'time' is similar to that used in
the adhesion model (Shandarin \& Zel'dovich 1989).

As was noticed in DD99 the structure evolution shows strong
features of self-similarity and is described by universal
expressions depending on the dimensionless variables ${\bf q}/l_v$
and $\tau$. This is a direct consequence of the Zel'dovich
approximation.

The `time' $\tau$ can be measured by different methods,
some of which are discussed below. It is sensitive to the
sample under investigation and to the method of measurement.
It can be used to quantify bias between spatial distributions
of different objects, such as, for example, large scale bias
between distributions of galaxies and the DM component.

The quadrupole component of the CMB
anisotropy, $T_Q$, the variance of density in a sphere with
radius $8h^{-1}$Mpc, $\sigma_8$, and the velocity dispersion,
$\sigma_{vel}$ are the more often used characteristics of the large
scale amplitude. All these characteristics are proportional
to each other, but their dependence on $\Omega_m$ and $h$ is
different, and they are sensitive to matter distribution in
different scales. Thus, the quadrupole component of CMB
anisotropy characterizes the perturbations on scales comparable
with the horizon, while the values $\sigma_{vel}$ and $\sigma_8$
are more sensitive to the matter distribution in moderate and
small scales.

The connection of these characteristics with $\sigma_s$ and
$\tau$ can be summarized as follows:
\begin{enumerate}
\item{}
Using the fits for the CMB anisotropy proposed by Bunn~\&~White
(1997) we obtain for the flat $\Lambda$CDM and open OCDM models
$$\tau_T\approx 2.73~h^2\Omega_m^{1.2}\left({m_{-2}\over 0.023}
{T_Q\over 20\mu K}\right),~\Omega_\Lambda=1-\Omega_m,   \eqno(2.9)$$
$$\tau_T\approx 2.73~h^2\Omega_m^{1.65-0.19\ln\Omega_m}\left({m_{-2}
\over 0.023} {T_Q\over 20\mu K}\right),~ \Omega_\Lambda=0,$$
where $\tau_T$ denotes the amplitude of large scale perturbations,
$\tau$, measured by the CMB anisotropy. These estimates depend 
on the spectral moment $m_{-2}$ only which is very stable and 
does not change during the considered period of evolution. But 
the estimates should be corrected if a possible contribution 
of gravitational waves is taken into account.

\item{} The amplitude of perturbations, $\sigma_s$ and $\tau$,
can be directly expressed through the two point autocorrelation
function as follows:
$$\sigma_s^2 = \lim_{r\rightarrow\infty}
\int_0^r dx \left(1-{x\over r}\right)x\xi(x),\eqno(2.10)$$
and for the autocorrelation function $\xi(r)$ approximated by the
power law
$$\xi(r) = (r_0/r)^\gamma,\quad r\leq r_\xi,\eqno(2.11)$$
 we have
$$\sigma_s^2(r_\xi)\approx {r_\xi^{2-\gamma}r_0^\gamma\over (2-\gamma)
(3-\gamma)},\quad \tau_\xi = {\sigma_s(r_\xi)\over\sqrt{3}l_v}.
						\eqno(2.12)$$
Here $r_\xi$ is the first zero-point of the autocorrelation 
function and $\tau_\xi$ denotes the amplitude $\tau$ measured 
by this function. The parameter $r_\xi$ is usually
found with small precision, but for $\gamma\approx$1.5 -- 1.7,
$1-\gamma/2\approx$0.25 -- 0.15 even some variations of
$r_\xi$ do not change significantly the final estimate of $\tau$.

\item{} The parameter $\sigma_8$ can be also expressed through
the two point autocorrelation function, $\xi(r)$, (Peebles 1993),
and for $\xi(r)$ approximated by a power law (2.11) we have
$$\sigma_8^2 ={72\over (3-\gamma)(4-\gamma)(6-\gamma)}
\left({r_0\over 16h^{-1}{\rm Mpc}}\right)^\gamma,\eqno(2.13)$$
$$\sigma_s^2\approx \sigma_8^2 (8h^{-1}{\rm Mpc})^2
{(4-\gamma)(6-\gamma)\over 18(2-\gamma)}
\left({r_\xi\over 16h^{-1}{\rm Mpc}}\right)^{2-\gamma},$$
$$\tau_8=\sigma_8\Gamma\sqrt{(4-\gamma)(6-\gamma)\over
36.75(2-\gamma)}\left({r_\xi\over 16h^{-1}{\rm Mpc}}
\right)^{2-\gamma\over 2}.			\eqno(2.14)$$
\item{} The dispersion of the peculiar velocity of particles at
small redshifts, z=0, can be written as in the linear theory
(DD99)
$$\sigma_{vel} = u_0\sqrt{3}\tau,~
u_0=l_vH_0\beta\approx {\Omega_m\over \Gamma}
{1535\mbox{km/s}\over 1+1.3\Omega_m},	 	\eqno(2.15)$$
and for $\tau$ we obtain the independent estimate
$$\tau_{vel} = {\sigma_{vel}\over \sqrt{3} u_0}. \eqno(2.16)$$
Here $\tau_{vel}$ denotes the amplitude $\tau$ measured by the
velocity dispersion. $\sigma_{vel}$ takes into account also the high
velocities generated by the gravitational compression of matter
(in particular, within clusters of galaxies) and, so, it gives
actually an upper limit of the amplitude.
\end{enumerate}

\section{Statistical characteristics of walls in
the Zel'dovich theory}

In both observed and simulated catalogues, at small redshifts, 
the wall-like structure elements accumulate $\sim$ 50\% of galaxies
and form the skeleton of large scale structure. So, investigation 
of the characteristics of these elements is important in
itself. It allows us also to obtain information about processes 
of nonlinear structure evolution. In particular, we can find two 
independent measures of the large scale amplitude, $\tau$. As 
walls represent the first step of the large scale nonlinear 
matter compression their characteristics can be compared with 
predictions of the Zel'dovich theory.

In this Section we will consider five characteristics of walls, 
namely, the surface density of walls, $m_w$, defined as the amount
of matter per unit of wall surface, for example, per $h^{-2}$Mpc
$^2$, the thickness of walls, $h_w$, the wall separation,
$D_{sep}$, the velocity dispersion of matter compressed within
walls, $w_w$, and the dispersion of wall velocities, $\sigma_v$.
All these characteristics can be derived from  the Zel'dovich theory
(DD99) and can be found for simulated point distributions as well. 

\subsection{Formation of walls}

Following DD99, we will consider the intersection of two fluid
particles with Lagrangian coordinates ${\bf q}_1$ and ${\bf q}_2$
as the formation of a wall (Zel'dovich pancake) with the surface
density $m_w=\langle n_p\rangle |{\bf q}_1-{\bf q}_2|$
where $\langle n_p\rangle$ is the mean particles density in the
sample. In Zel'dovich theory statistical characteristics of such
walls are described by the initial power spectrum (2.5)
and can be expressed through the characteristic scale, $l_v$,
the surface density of wall, $m_w$, or dimensionless surface
density, $q_w=m_w/l_v/\langle n_p\rangle$, and the `time',
$\tau$, introduced in Secs. 2. To do this, the structure functions
of the initial power spectrum can be used. For the standard SDM-like
power spectrum (2.5) with the BBKS transfer function these functions
were introduced in DD99.

Naturally, the theoretical considerations describe the idealized
model of structure evolution. Thus, it uses the rigid wall boundary
though in reality such boundaries are always blured. Other
important factor is the compression and expansion of pancakes
in transversal directions. These motions transform pancakes into
filaments and/or lead to the dissipation of poor pancakes. They
are not so important for rich walls but can change the wall
surface density by a factor of 1.3 -- 1.5. The small scale
clustering and relaxation of matter distorts also the measured
characteristics of walls with respect to theoretical expectations.

These factors distort the actual power spectrum with respect to the used
one and introduce differences between the expected
and actually measured parameters of walls which however cannot
be evaluated {\it a priori}. The actual power and limitations
of this approach must be tested at first with N-body simulations.

\subsection{Wall properties in the real space}

\subsubsection{Surface density of walls}

The most fundamental characteristic of walls is the surface
density, $m_w$. The approximate expression for the probability
distribution function (PDF) of the pancakes surface density, $m_w$,
defined as above has been obtained in DD99 in the
same manner as the well known Press-Schechter mass function. 
It characterizes the process of one dimensional matter compression
and formation of wall-like pancakes as described by the Zel'dovich
theory.

For Gaussian initial perturbations and the standard CDM-like
power spectrum with the BBKS transfer function, it can be
written as follows:
$$N_m = {1\over \sqrt{2\pi}\tau_m}{1\over\sqrt{q_w}}
\exp\left(-{q_w\over 8\tau_m^2}\right) \mbox{erf}\left(\sqrt{q_w
\over 8\tau_m^2}\right),\eqno(3.1)$$
$$q_w = {m_w\over l_v\langle n_p\rangle}={|{\bf q}_1-{\bf q}_2|
\over l_v},\quad \int_0^\infty N_m(q_w)~dq_w = 1,$$
$$\langle q_w\rangle =\int_0^\infty q_w N_m(q_w)~dq_w = 
8(0.5+1/\pi)\tau_m^2\approx 6.55\tau_m^2,$$
$$\langle q_w^2\rangle = \int_0^\infty q_w^2 N_m(q_w)~dq_w = 
128(0.375+1/\pi)\tau_m^4\approx 887\tau_m^4, $$
where $\langle n_p\rangle$ is the mean particle density in the
sample, $l_v$ is defined by (2.7), $\tau_m$ characterizes the
amplitude of perturbations and the evolution stage of structures,
$\tau$, as measured by the surface density of walls, and
${\bf q}_1~\&~{\bf q}_2$ are Lagrangian coordinates of wall
boundaries. This relation was corrected for the merging of
neighboring walls and this process is described by the erf -
function in (3.1).

These expressions connect $\tau_m$ with the mean surface density
of walls and allow us to estimate $\tau_m$ from measurements of
$\langle q_w\rangle$. For other models and/or other distributions
of initial perturbations the PDFs similar to (3.1) could be
obtained using the technique described in DD99.

\subsubsection{The wall separation}

We have not been able to find a simple theoretical description of 
the wall separation. Nonetheless, taking into account the mainly 
one dimensional character of wall formation, we can roughly link
the mean measured wall separation, $\langle D_{sep}\rangle$,
to the mean surface density of walls, $\langle q_w\rangle$.

Indeed, the matter conservation law along the direction
of wall compression can be approximately written as follows:
$$\langle m_w\rangle\approx f_w\langle n_p\rangle\langle
D_{sep}\rangle,\quad \langle q_w\rangle\approx f_w\langle
D_{sep}\rangle/l_v,$$
where $f_w$ is the matter fraction assigned to walls. It
implies that on average a fraction $f_w$ of particles situated
at the distance $\pm 0.5D_{sep}$ from the center of wall will
be collected by the wall. For simulations when the mean wall
separation is comparable to the box size, $L_{box}$, we will
use the more accurate relation
$$\langle q_w\rangle\approx {f_{dq}\over l_v}\left\langle
{D_{sep}\over 1+D_{sep}/L_{box}}\right\rangle.\eqno(3.2)$$
The averaging can be performed analytically assuming the exponential
distribution function for the wall separation.

The factor $f_{dq}$ defined by Eq. (3.2) characterizes the matter
fraction assigned to walls as it is determined by comparison of
independently measured characteristics $\langle q_w\rangle$ and
$\langle D_{sep}\rangle$. In turn, difference between
$f_{dq}$ and $f_w$ characterizes the robustness and degree of self
consistency of the model and the measurements. These estimates are
only approximations because the wall formation is actually a
three dimensional process.

\subsubsection{Velocity of structure elements}

For the pancakes defined  in Sec. 3.1 the 1D velocity of walls,
$v_w$, can be found from relations (2.2) and (2.3) as follows:
$$v_w = {1\over |{\bf q}_1-{\bf q}_2|}\int_{q_1}^{q_2} {\bf n}
[{\bf u}-H(z){\bf r}]~dq\eqno(3.3)$$
where ${\bf n}$ is a unit vector normal to the wall. The small
scale clustering and relaxation of compressed matter does
not influence the velocities of walls and, so, they are the most
stable characteristics of the evolutionary stage reached. As was
shown in DD99 the mean velocity of walls, $\langle v_w\rangle$,
is expected to be negligible as compared with its dispersions,
$\sigma_{v}$, and the expected PDF of this velocity, $N_v$, is
Gaussian for Gaussian initial perturbations.

For the standard spectrum (2.5) with the BBKS transfer function,
and for $q_w\ll$ 1, the velocity dispersion is related to the
amplitude of initial perturbations as follows:
$$\sigma_{v}\approx u_0\tau,\quad
\tau_{v} = {\sigma_{v}\over u_0},\eqno(3.4)$$
what is similar to (2.16) and also is identical to expectations of
the linear theory. Here $\tau_v$ denotes the amplitude $\tau$ as
measured by the dispersion of wall velocity and $u_0$ was introduced
by (2.15).

\subsubsection{Velocity dispersion of matter compressed
within walls}

The variance of velocity of matter accumulated by walls,
$$w_{wz}^2 = {1\over |{\bf q}_1-{\bf q}_2|}\int_{q_1}^{q_2}
[{\bf n}{\bf u}-H(z){\bf n}{\bf r} - v_w]^2dq,	    \eqno(3.5)$$
can be found in the framework of the Zel'dovich theory
using the structure functions described in DD99. As is shown in
Appendix A, it can be written as follows:
$$w_{wz}^2(q_w,\tau)\approx u_0^2\left({q^2_w\over 12}+
{\tau^2(1+\beta)^2\over 3\beta^2}q_w\right),~~ q_w\ll 1,\eqno(3.6)$$
where $\beta,~u_0,~q_w~\&~v_w$ were introduced by (2.3), (2.4),
(2.15), (3.1) \& (3.3). In fact, this function characterizes the
mean kinetic energy of particles compressed into a wall of a
given size $q_w$. After averaging over a sample of walls with
the PDF $N_m$ (3.1), in the Zel'dovich theory, we obtain
$$ w_z^2(\tau)=\langle w_{wz}^2(q_w,\tau)\rangle \approx
u_0^2\tau^4\left(7.4 +{2.2(1+\beta)^2\over \beta^2}
\right).  						\eqno(3.7)$$
The comparison of the expected mean kinetic energy of the compressed 
particles with the kinetic energy measured in simulations characterizes 
the mean degree of relaxation of compressed matter at a given $\tau$.

For richer walls, with $q_w\gg\langle q_w\rangle$, the relation
(3.6) is transformed into
$$w_{wz}\approx {u_0\over\sqrt{12} }q_w,    	\eqno(3.8)$$
and for such walls, the PDF is similar to (3.1).
For a rich sample of walls, this relation can be also used for
the direct measurement of the amplitude $\tau$ (DMRT; DD99).

\subsubsection{Wall thickness}

The methods discussed in DD99 allow us also, in the framework
of the Zel'dovich theory, to obtain the expected thickness of
walls along the direction of maximal compression, $h_w$. It
can be characterized by the thickness of a homogeneous slice
with the same surface density. The corresponding expression
(Appendix A) is
$$h_{wz}(q_w,\tau)\approx 2l_v\tau \sqrt{q_w}~(1+z)^{-1}. \eqno(3.9)$$
This relation shows that the wall thickness is strongly correlated
with its surface density. After averaging with the PDF (3.1) we
obtain for the mean thickness of walls
$$\langle h_{wz}\rangle\approx 8\pi^{-1/2} l_v\tau^2(1+z)^{-1}.
						\eqno(3.10)$$

The degree of matter compression in the Zel'dovich theory,
$\delta_z(q,\tau)$, is characterized by the ratio
$$\delta_z = {q_wl_v\over h_w} = {\sqrt{q_w}\over 2\tau}.$$
After averaging with the PDF (3.1) we have for the mean degree
of matter compression
$$\langle \delta_z\rangle\approx {\langle \sqrt{q_w}\rangle\over
2\tau}={2\over \sqrt{\pi} } = 1.13.\eqno(3.11)$$
So, in the Zel'dovich theory  the averaged degree of matter compression
is small.

\subsection{Wall properties in the redshift space}

In observed catalogues only the redshift position of galaxies
along the line-of-sight is known, and therefore the parameters
of observed structures with respect to those found above can
differ due to the influence of the velocity field. The statistical
characteristics of walls in redshift space predicted by the
Zel'dovich theory can be found with the methods described above.
This information is not so rich as in the real space because in
the redshift space, positions of particles are determined by
their velocities, and, for example, such a useful characteristic
as the wall velocity cannot be found.

\subsubsection{Surface density of walls}

In the real space (Sec. 3.1) the pancake formation was defined
as an intersection of particles with coordinates ${\bf q}_1~
\&~{\bf q}_2$. In the redshift space the velocity (2.3) along
the line-of-sight must be used instead of the coordinate. In
Zel'dovich theory the velocity dispersion exceeds the
dispersion of displacement by a factor of $(1+\beta)$. Hence,
this substitution increases the wall surface density in the
redshift space in respect to that in the real space, and now
we must use
$$\tau_{rd} = f_{rd}\tau = \tau\sqrt{(1+\beta)^2\cos\phi^2 +
\sin\phi^2},					\eqno(3.12)$$
instead of $\tau$. Here the factor $f_{rd}\geq$1 describes the
more effective matter compression in the redshift space predicted
by the Zel'dovich theory, $\beta$ was introduced in (2.3),
(2.4), and $\phi$ is a random angle between the direction
of wall compression and the line-of-sight ($0\leq\phi\leq\pi/2$).
Evidently, $\tau_{rd}=\tau$ for $\beta=0$, so $f_{rd}(\beta=0)=$1.

The PDF of wall surface densities in the redshift space is
identical to (3.1) with a substitution of $\tau_{rd}$ for
$\tau$, and now for the mean surface density of walls we have
$$\langle q_w\rangle =8(0.5+1/\pi)\langle f_{rd}^2\rangle
\tau^2\approx 6.55\langle f_{rd}^2\rangle\tau^2,\eqno(3.13)$$
$$1\leq\langle f_{rd}^2\rangle ={1\over 3}[2+(\beta+1)^2]\leq
3.667,$$
where $\tau$ characterizes the evolutionary stage as before.
Probably, these relations can be used for the description of
poorer pancakes and earlier evolutionary stages when the influence
of other factors is less important.

At small redshifts we must take into account the influence of
the high velocity dispersion of compressed matter generated
by the small scale matter clustering and relaxation. The
influence of this factor, well known as the `finger of God'
effect, is opposite to that discussed above. It changes
the observed particle position within walls along the line-of-sight
what blurs the wall boundary and increases the thickness
of observed walls. It artificially removes the high velocity
particles from the selected wall and effectively decreases
the surface density of walls selected in redshift space with
respect to the estimates (3.13).

The impact of this factor can be approximately described by a
modification of PDF of wall surface density, 
$$N_m^{rd} = {1\over \sqrt{2\pi}f_{rd}\tau}{1\over
\sqrt{q_w}} \mbox{erf}\left(\sqrt{q_w\over 8f_{rd}^2\tau^2}
\right)\times 				  \eqno(3.14)$$
$$\left[\exp\left(-{q_w\over 8f_{rd}^2\tau^2}\right)-\exp\left(-
{q_w\over 8\tau^2}\right)\cdot W(q_w,\tau,\delta_{thr})\right],$$
and a new normalization of distribution $N_m^{rd}$.

The second term in the square brackets describes the artificial
rejection of high velocity particles from the wall with a
surface density $q_w$ bounded by a threshold density $\delta_{thr}$.
In this term the exponent gives the fraction of matter
accumulated by the wall in real space for some $q_w~\&~\tau$,
whereas $W(q_w,\tau,\delta_{thr})$ is the fraction of high velocity
particles which are removed from the wall in the redshift space.
The function $W(q_w,\tau,\delta_{thr})$ cannot be found in the
Zel'dovich theory as it depends on the distributions of
particles positions and velocities arising due to the small
scale clustering and relaxation of matter compressed within
walls.

An other factor which can suppress the expected difference of wall
characteristics, measured in the real and redshift spaces at small
redshifts, is the strong matter condensation within structure elements
with various richnesses.  The strong matter rearrangement transforms the
continuous matter infall on walls into a discontinuous one, increases
the separation of infalling structure elements, even in the redshift
space and, so, at least partly, prevents the erosion of wall boundaries.

These comments show that in the redshift space the Zel'dovich theory 
with the factor $f_{rd}$ given by (3.12) ~\&~ (3.13) overestimates 
the matter concentration within walls. Therefore, instead of
the factor $f_{rd}$ in (3.13) a factor $\kappa_{rd}(\Gamma,
\tau,l_{thr})$ should be used and the more realistic relation
$$\tau_m\approx \sqrt{q_w\over 6.55\kappa_{rd}^2},\quad
1\leq \kappa_{rd}\leq f_{rd},\eqno(3.15)$$
connects the amplitude $\tau_m$ with the wall richness
$q_w$ in the redshift space.

The actual value of $\kappa_{rd}$ depends on the parameters of
the cosmological model and on the method of identification of walls.
The analysis performed below shows that for the walls selected
in 3D space as described in Sec. 6.1, no growth of $q_w$ was
found, and the parameters $q_w~\&~ \tau_m$ are connected by the
relation (3.1) as in the real space.

\subsubsection{Wall separation}

The separations of richer walls is not sensitive to relatively
small shifts of particle positions introduced by the random
velocities, but these shifts can result in an artificial merging
of poorer walls. The influence of this factor can be tested 
with the relation (3.2) as before.

\subsubsection{Velocity dispersion of matter compressed
within walls and the wall thickness}

In the redshift space the expression for the velocity dispersion
of matter compressed within walls in the Zel'dovich theory is
identical to (3.6) with a substitution of $\tau_{rd} =\tau
\kappa_{rd}$ instead of $\tau$, but now it
characterizes also the observed thickness of walls. For
walls selected from the 3D sample of particles, as is
described in Sec. 6.1, we have
$$h_w = \sqrt{12}w_wH_0^{-1}, 	\eqno(3.16)$$
$$w_w=u_0\sqrt{{q_w^2\beta^2\over 12}+
{\tau^2\kappa_{rd}^2q_w\over 3}(1+\beta)^2}.	\eqno(3.17)$$
This value exceeds the corresponding real thickness of walls
given by (3.9). The expected overdensity of compressed matter
is given by
$$\delta_{rd} = l_v\langle q_w/h_w\rangle.	\eqno(3.18)$$

 \section{Measured characteristics of large scale matter
distribution.}

\subsection{Core-sampling approach}

The core-sampling approach was proposed by Buryak et al. (1994)
for the analysis of the galaxy distribution in deep pencil beam
redshift surveys. In the original form it allows to obtain the
mean free-path between the filaments and walls. It was improved
and described in detail in LCRS1 where some characteristics of
the large scale galaxy distribution were found for the Las
Campanas Redshift Survey. For simulated matter distributions
as considered here these characteristics were discussed in DMRT.

The potential of the core-sampling approach is not exhausted by
these applications, and it could be used to measure
parameters of the large scale matter distribution discussed in the previous
Sections. Here we will use this approach to obtain the characteristics
of the wall-like structure component.

The core-sampling method deals with a sample of points
(galaxies) lying within relatively narrow cores -- rectangular and/or
cylindrical in simulations, and conical in observations -- and it
studies the point distribution along these cores. For some
applications the transversal coordinates of points can be used as
well. To take into account the selection effects, which are
important for observed catalogues, appropriate corrections can
be incorporated. The sampling core is characterized by the size, $D_{core}$,
that is the side of a rectangular core or the angular diameter of a
conical core.

\subsection{Measured characteristics of walls}

Here we will apply the core sampling technique to the sample of
wall-like structure elements selected by a 3D-cluster analysis
(DMRT, Sec. 6.1). This means, the sampling cores contain only the
particles assigned to walls. Further on, all particles are projected
onto the core axes and are collected into a set of clusters with a
linking length $l_{link}$. Clusters with richness larger than a
threshold richness, $N_{min}$, are identified with walls within
the sampling core.

The measured wall parameters are sensitive to the influence of
small scale clustering of matter within walls. For strongly
disrupted walls and a narrow core, the results depend on the random
position of high density clumps, what strongly increases
the scatter of measured wall properties. The influence of this
factor is partly suppressed for larger sizes of the sampling core,
$D_{core}$.

However, the random intersection of the core 
with a wall boundary generates artificially poor clusters. The number
of such intersections increases proportionally to $D_{core}$ what
restricts the maximal $D_{core}$. To suppress the influence of this
factor a threshold richness of cluster, $N_{min}$, was used. If
however $N_{min}$ becomes too large, the statistical estimates become
unreliable. For large $D_{core}$ the overlapping of projections
of neighboring walls becomes also important what distorts the measured
wall characteristics.

It is also important to choose an optimal linking length, $l_{link}$,
because for small $l_{link}$, only the high density part of walls
is measured, whereas for larger $l_{link}$, again the impact of the
random overlapping of wall projections becomes important.

The influence of these factors cannot be eliminated completely,
and our final estimates of properties of walls are always
distorted to some degree. These distortions can be minimized
for an optimal range of parameters $D_{core}, N_{min} ~\&~l_{link}$.
Practically, these factors do not distort the velocity dispersion
of walls, $\sigma_v$, which therefore provides the best characteristic
of the actual evolutionary stage of the wall formation. On the
other hand, the comparison of results obtained for different
$l_{link}$ and $D_{core}$ allows to characterize the inner structure
of walls.

\subsubsection{Measurement and correction of wall parameters}

The richness of clusters in the core measures the surface density
of walls,
$$m_{sim} = {N_m\over D_{core}^2},\eqno(4.1)$$
where $N_m$ is the number of particles in a cluster. The velocity
of walls, $v_{sim}$, the velocity dispersion of particles
accumulated within walls, $w_{sim}$, and the proper sizes of
walls, $h_{sim}$, are found as follows:
$$r_w = {1\over N_m}\sum_{i=1}^{N_m}r_i,~~
v_{sim} = {1\over N_m}\sum_{i=1}^{N_m}(u_i-Hr_i),$$
$$w_{sim}^2 = {1\over N_m-1}\sum_{i=1}^{N_m}(u_i-Hr_i-v_{sim})^2,
\eqno(4.2)$$
$$ h_{sim}^2 = {12\over N_m-1}\sum_{i=1}^{N_m}(r_i-r_w)^2.$$
Here $r_i$, $r_w$ and $u_i$  are the coordinates of a particle,
of a wall, and the velocity of a particle along the sampling core,
respectively. The wall separation, $D_{sim}$, is measured by the
distance between neighboring clusters.

The parameters $m_{sim}$, $v_{sim}$, $w_{sim}$ and $h_{sim}$ as
given by (4.1) and (4.2) are found along the sampling core and,
so, are not identical to the parameters discussed in Sec. 3.
These parameters must be corrected for the random orientation
of walls with respect to the  sampling core. The impact of this
factor increases the measured surface density, and the corrected
wall surface
density, $m_c$, is connected with the measured one by
$$m_c = m_{sim}\cos\phi,\quad 0\leq \phi\leq\pi/2,$$
$$\langle m_c\rangle = 0.5\langle m_{sim}\rangle,     \eqno(4.3)$$
where $\phi$ is a random polar angle between the core and the
vector orthogonal to the surface of the wall, and the averaging
is performed in a spherical coordinate system. Corrected values
of the wall velocity and the walls thickness are as
follows:
$$v_c = v_{sim}\sqrt{3},\quad h_c=h_{sim}/\sqrt{3}.    \eqno(4.4)$$
In the redshift space the wall thickness is connected with the velocity
dispersion by Eq. (3.16). The velocity dispersion within walls
was found to be almost isotropic (DMRT), and, so, we will use
the measured $w_{sim}$ as the actual velocity dispersion across
walls.

The measured PDF of the wall surface density, $N_m(m_c)$, and
the mean wall surface density, $\langle m_c\rangle$, are
distorted due to the small statistics of rich walls and
rejection of poor walls with a richness $N_m\leq N_{min}$.
The correction for these distortions can be estimated by
comparing the simulated PDF with the expected PDF (3.1).

To do this we will fit the measured PDF to the function
$$N_m = {a_m\over \sqrt{x_m}}e^{-x_m}\mbox{erf}(\sqrt{x_m}),
\quad x_m={b_m m_{sim}\over \langle m_{sim}\rangle}. \eqno(4.5)$$
The parameter $b_m$ describes deviations of measured and
expected mean surface density of walls $\langle m_{sim}\rangle$,
and $a_m$ is a normalization factor. If the measured PDF is
well fitted to the function (4.5) then the value
$$m_t =\langle m_c\rangle/b_m		\eqno(4.6)$$
can be taken as a measure of the `true' mean surface
density of walls.

\begin{table*}
\begin{minipage}{120mm}
\caption{Parameters of simulated DM and mock catalogues}
\label{tbl1}
\begin{tabular}{ccc ccc ccc c} 
sample &$\Omega_m$&h&$\tau_T$&$\sigma_8$&$r_0$&$\gamma$&$\tau_\xi$&
$\sigma_{vel}/\sqrt{3}$&$\tau_{vel}$\cr
       &      &   & & &$h^{-1}$Mpc&      &    &km/s&   \cr
SCDM        &1~~~~ &0.5&0.68&1.37&6.5&1.9&0.94&670&0.51\cr
$\Lambda$CDM&0.35  &0.7&0.37& 1.11&6.0&1.8&0.34&554&0.37\cr
OCDM        &0.5~~ &0.6&0.29& 0.74&5.0&1.3&0.25&346&0.23\cr
$mock_1$    &0.5~~ &0.6&0.29& 0.95&6.0&1.4&0.28&370&0.25\cr
$mock_2$    &0.5~~ &0.6&0.29& 0.95&6.0&1.4&0.28&370&0.24\cr
$mock_3$    &0.5~~ &0.6&0.29& 1.24&7.0&1.5&0.33&374&0.24\cr
$mock_4$    &0.5~~ &0.6&0.29& 1.61&8.0&1.6&0.39&404&0.26\cr
\hline
\end{tabular}

$r_0 ~\&~ \gamma$ are the correlation lengths and the slope
of correlation function (2.11) in redshift space; $\sigma_8$,
$\tau_T$, and $\tau_\xi$ are the amplitudes of perturbations as
given by (2.9) ~\&~ (2.12), $\sigma_{vel}~\& ~\tau_{vel}$ are
the velocity dispersion of all particles and amplitudes of
perturbations measured by $\sigma_{vel}$ as given by (2.16);
\end{minipage}
\end{table*}

Finally, the mean dimensionless surface density of walls,
$\langle q_w\rangle$ and the amplitudes of perturbations,
$\tau_m~\&~\tau_v$, measured by the surface density and velocity
of wall-like structure elements, can be estimated as follows:
$$\langle q_w\rangle = {\langle m_{sim}\rangle\over 2b_ml_v
\langle n_p\rangle}, ~\tau_m= \sqrt{\langle q_w\rangle
\over 6.55},~ \tau_{v} = {\sqrt{\langle v_{sim}^2\rangle}
\over u_0}.\eqno(4.7)$$

The small statistics of rich  and poor  walls distorts also
the measured wall separation, $D_{sep}$. The expected distribution
of wall separations is exponential, and therefore it is possible
to correct the mean separation using the fit of the measured
PDF, $N_{sep}(D_{sim})$, to the function
$$N_{sep}=a_{sep}\exp(-b_{sep}D_{sim}/\langle D_{sim}\rangle).
						\eqno(4.8)$$
As before, the parameter $b_{sep}$ describes deviations of the
measured and expected mean separation of walls, and $a_{sep}$
is a normalization factor. If the measured PDF is well fitted
to the function (4.8) then the value
$$\langle D_{sep}\rangle =\langle D_{sim}\rangle/b_{sep},
						    \eqno(4.9)$$
can be taken as a measure of the `true' mean separation of walls.

\section{General characteristics of the simulated matter distribution}

\subsection{Basic simulations}

The theoretical model discussed above describes the evolution
of the DM distribution and, so, should be tested with the simulated
DM distribution as well. Here we use three simulations as a
basis for our analysis -- the COBE normalized standard CDM model (SCDM),
a $\Lambda$CDM with $\Omega_{\Lambda}>\Omega_{m}$, and an open
CDM (OCDM) model. These models were described and investigated
with 3D cluster analysis and Minimal Spanning Tree technique
in DMRT. It was found that the $\Lambda$CDM and OCDM models
successfully reproduce
the main observed characteristics of large scale
matter distribution while the SCDM model demonstrates strong
signatures of overevolution. Here we study these three models
bearing in mind that only the $\Lambda$CDM and OCDM models can be
considered as realistic models of the observed large scale
matter distribution. The SCDM model represents the matter
distribution typical for a late evolutionary stage.

The simulations were performed with a PM code in a box of
(500$h^{-1}$Mpc)$^3$ with (300)$^3$ particles for the
Harrison-Zel'dovich primordial power spectrum and the BBKS
transfer function. The force and mass resolutions are
$\sim 0.9 h^{-1}$Mpc and $\sim 10^{11}M_\odot$,
respectively. The point distribution in redshift space
was produced by adding an apparent shift to one coordinate
due to the peculiar velocity of particles.

Four mock catalogues were prepared on the basis of the OCDM
model with various degrees of large scale bias between the
spatial DM distribution and the `galaxies'. These mock
catalogues were constructed by identifying randomly `galaxies'
with DM particles, but with a probability depending on the
environmental density, thereby identifying more particles
as `galaxies' in high density regions (walls). These
catalogues were investigated also both in real and redshift
spaces.

The main characteristics of the simulations are listed in
Table 1. A more detailed description can be found in DMRT.

\subsection{Large scale amplitude of perturbations}

The evolutionary stages reached in the models under discussion
can be suitably characterized using the methods described in
Sec. 2. The value $\tau_T$ listed in Table 1 characterizes
the large scale amplitude used for the normalization of
simulated perturbations.
Other measures of the amplitude, such as $\sigma_8, \tau_\xi
~\&~ \tau_{vel}$, are sensitive to both the actually realized
sample of random perturbations (cosmic variance) and to the
nonlinear distortions of power spectrum produced during the
evolution. For the considered mock catalogues these measures
are also sensitive to the large scale bias between the spatial
DM and `galaxies' distributions what allows us to characterize
it quantitatively.

The spatial matter distribution and the bias between spatial
distributions of DM component and 'galaxies' can be
characterized by the correlation length, $r_0$, and the
slop of the correlation function, $\gamma$, introduced
in (2.11). These parameters are listed in Table 1 for all
samples. Using relations (2.12) and (2.13), these
values allows to calculate  $\sigma_8$ and $\tau_\xi$,
which are also listed in Table 1.

The characteristics of correlation function, $r_0$ and
$\gamma$, are sensitive to the perturbations in scales
$k\sim$0.5 -- 0.1h Mpc$^{-1}$. As is seen from (2.12),
estimates $\tau_\xi$ are very sensitive to the value of
$2-\gamma$, and, so, to small scale perturbations. The first
zero--point of autocorrelation function, $r_\xi\approx
40h^{-1}$Mpc, can be usually found with a large uncertainty
($\sim$20 -- 30 per cent) but its impact is reduced by the
small exponent $1-\gamma/2\leq 0.3$ in (2.12).

For OCDM and $\Lambda$CDM models the impact of small scale
matter clustering is moderate, and differences between $\tau_
\xi$ and $\tau_{vel}$ are found to be $\sim$10 per cent.
The differences between the same parameters and $\tau_T$
can be considered as a reasonable measure of simulated `cosmic
variance'. For these models differences between the parameter
$\tau_\xi$ calculated for the real and redshift spaces also do not
exceed $\sim$ 10\%. For the SCDM model both $\tau_\xi$ and
$\tau_{vel}$ are distorted by the strong small scale clustering.
This divergence indicates that for the SCDM model the successful
application of methods discussed in Sec. 3 is also in question.

The progressive growth of $\tau_\xi$ and $\sigma_8$ for mock
catalogues characterizes the degree of the large scale bias
between the spatial distribution of DM component and `galaxies'.

\section{Properties of wall -- like structure elements}

The main basic characteristics of walls were discussed in DMRT
for three DM and four mock catalogues mentioned above both in the
real and redshift spaces. In this Sec. the wall characteristics
discussed in Sec. 3 are found with the core-sampling technique for
the same simulations and the same samples of walls.

\subsection{Selection of wall-like structure elements}

The sample of wall-like structure elements was selected with the
two-parameter method described and exploited in DMRT. It
identifies the wall-like structure elements with
clusters found using a threshold linking length, $l_{thr}$,
and a threshold richness, $N_{thr}$. As usual, the boundary
of the clusters is defined by the threshold overdensity,
$\delta_{thr}$, which is connected with the threshold linking
length by
$$\delta_{thr} = {n_{thr}\over \langle n\rangle} = {3\over
4\pi\langle n\rangle l_{thr}^3}.\eqno(6.1)$$
The threshold richness, $N_{thr}$, restricts the matter
fraction, $f_w$, associated with walls.

The main characteristics of these samples both in real and
redshift spaces are listed in Table 2. The values of $f_w\approx$
0.4 -- 0.45 are consistent with the theoretically expected and
observed matter fraction accumulated by walls (DD99, LCRS1, LCRS2).
The analysis performed in DMRT shows that for the low density
models the main characteristics of such wall-like elements
are similar to the observed characteristics of superclusters
of galaxies (Oort 1983 a,b; LCRS2; DURS).

\subsection{DM walls in the real space}

The analysis of DM catalogues in the real space is most
interesting as in this case we can study the clear signal
from the gravitational interaction of compressed matter and
can reveal and characterize statistically the matter relaxation.
Five basic characteristics of DM walls discussed in Sec. 3,
namely, the wall thickness, $h_w$, the dispersions of wall
velocities, $\sigma_v$, the velocity dispersion of matter
compressed within walls, $w_w$, the dimensionless surface
density, $q_w$, and mean separation of walls, $D_{sep}$, can
be found with the core-sampling method and can be compared
with those found in DMRT. The surface density of walls is
closely connected with the size of proto-walls as discussed
in DMRT.

\begin{figure}
\centering
\epsfxsize=7.5 cm
\epsfbox{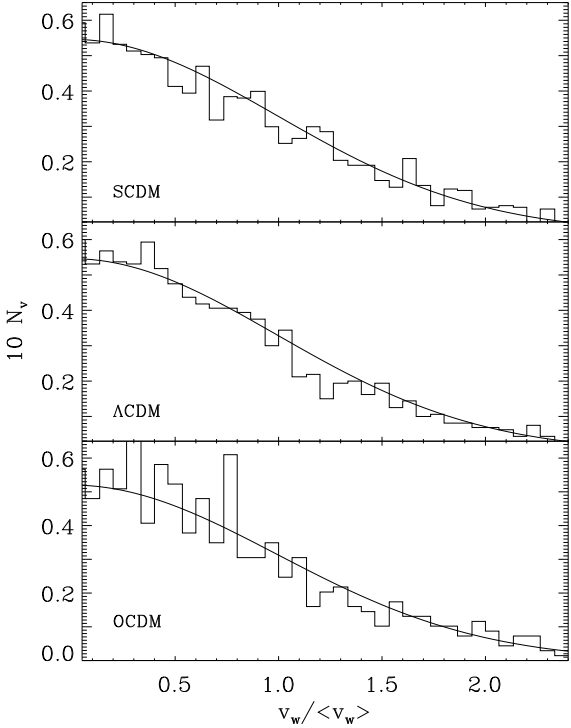}
\vspace{0.75cm}
\caption{PDFs of  DM wall velocity, $N_v(v_w/\langle v_w\rangle)$,
in real space for SCDM, $\Lambda$CDM and OCDM
models. The Gaussian fits are shown by solid lines.
}
\end{figure}

Comparison of such characteristics of matter distribution as
$\tau_{vel}$ listed in Table 1 and $\tau_v$ and  $\tau_m$ related
to the wall properties allows us to test the influence of small
scale matter clustering and other random factors discussed in
Sec. 4.3, and to find the optimal ranges of core size, $D_{core}$,
and of threshold richness, $N_{min}$, as well as the optimal
linking length, $l_{link}$. The results listed in Table 2 are obtained
with the linking length $l_{link} = 5h^{-1}$Mpc, and are averaged
over 7 core sizes, $6h^{-1}$Mpc$\leq D_{core}\leq 9h^{-1}$Mpc, and
over 7 threshold richness, 10$\leq N_{min}\leq$35 .

\subsubsection{Basic characteristics of DM walls}

For all models, the dispersion of wall velocities, $\sigma_v$,
is found to be the best and most stable characteristic
of the evolutionary stage reached. This is the
direct consequence of the discrimination between the wall
velocity and the velocity dispersion of particles compressed
within walls. The PDFs, $N_v$, plotted in Fig. 1, are well
fitted to Gaussian functions with the measured dispersion.

For the OCDM and $\Lambda$CDM models the mean dimensionless surface
density of walls, $\langle q_w\rangle$, and the amplitudes, $\tau_m
\approx\tau_v\approx\tau_{vel}$, are found with scatters $\sim$10
-- 15\% for the used $N_{min}$, $D_{core}$, and $l_{link}$. This
scatter characterizes the moderate action of random factors
discussed in Sec. 3.2 and the procedure of measurement. The values of
$l_v \langle q_w\rangle$ are consistent with estimates of the size
of proto walls obtained in DMRT. The PDFs of the surface density
plotted in Fig. 2 are consistent with that expected form (3.1). These
results demonstrate that for lower density cosmological models the
Zel'dovich approximation successfully describes these basic
characteristics of rich walls.

\begin{figure}
\centering
\epsfxsize=7.5 cm
\epsfbox{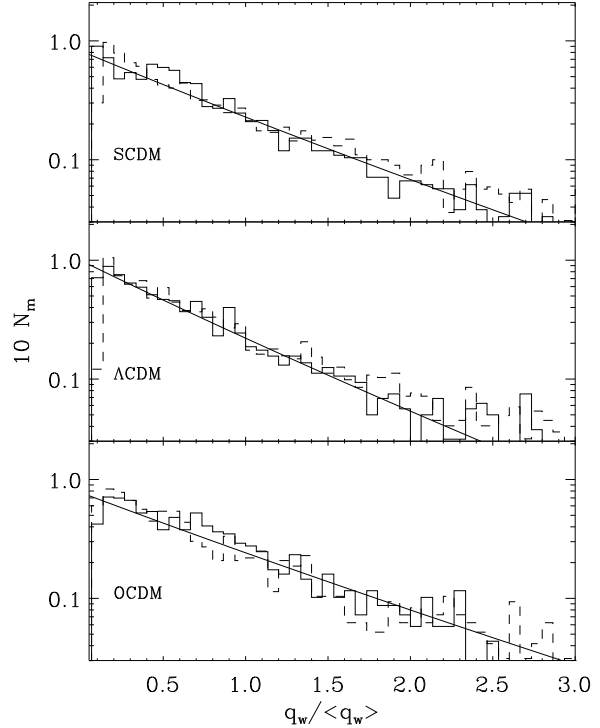}
\vspace{0.75cm}
\caption{PDFs of the DM  wall surface density, $N_m(q_w/\langle
q_w\rangle)$, in real (solid lines) and redshift (dashed lines)
spaces for SCDM, $\Lambda$CDM and OCDM models. The fits (3.1)
are shown by solid lines.
}
\end{figure}

For the SCDM model, the results listed in Table 2 are more sensitive
to the method of measurement and the surface density of walls is
underestimated, $\tau_m <~\tau_v\approx\tau_{vel}$.
This difference can be mainly ascribed to the strong
disruption of walls occurring at late evolutionary stages in
this model. Other important factors are the faster compression
and/or expansion of walls in transversal directions and the existence
of richer halos of evaporated particles around the walls mixed
with infalling particles. Such a halo becomes richer for larger
$\tau$, i.e. for the $\Lambda$CDM, and especially, for the SCDM
models.

The distribution function of wall separation, $N_{sep}$, plotted
in Fig. 3 is well fitted to (truncated) exponential distribution.
The mean  wall separation $\langle D_{sep}\rangle$ is sensitive
to the threshold richness $N_{min}$ and to the core size 
$D_{core}$. The separation $\langle D_{sep}\rangle\sim 40h^{-1}$Mpc,
found for the lower threshold richness, $N_{min}$=5, and larger
core sizes, $D_{core}=9h^{-1}$Mpc, coincides with the results obtained in
DMRT. It increases with $N_{min}$ as the number of rich walls
progressively decreases. For smaller $D_{core}$ and larger $N_{min}$
some of highly disrupted walls are lost due to their small covering
factor. This parameter can be found with relatively large scatter.

Using relation (3.2) we can compare our estimates of $\langle
D_{sep}\rangle$ and $\langle q_w\rangle$. For all models we have
$$f_{dq}\approx (0.75 - 0.9)f_w$$
and the mean wall separation is probably overestimated.

For all models under consideration, the mean wall thickness,
$\langle h_w\rangle$, is similar to that found in DMRT with
the inertia tensor technique, where a wall is represented by
a homogeneous ellipsoid. It is about of 2 -- 4 times smaller
than that expected in the Zel'dovich approximation (3.10)
what bears a sign of the relaxation of gravitationally bounded
DM particles within walls.

\begin{figure}
\centering
\epsfxsize=7.5 cm
\epsfbox{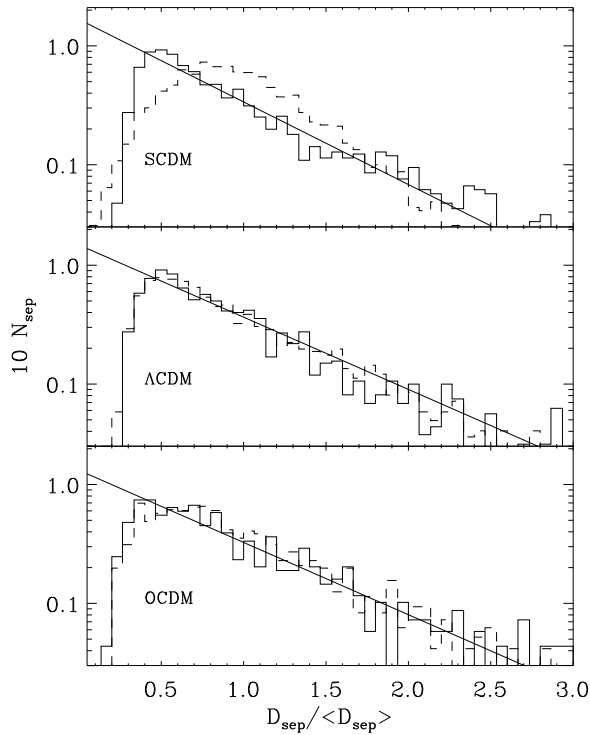}
\vspace{0.75cm}
\caption{PDFs of DM wall separations, $N_{sep}(D_{sep}/\langle
D_{sep}\rangle)$, in real (solid lines) and redshift (dashed
lines) spaces for SCDM, $\Lambda$CDM and OCDM models. The
exponential fits are shown by solid lines.
}
\end{figure}

For the OCDM model the velocity dispersion of matter compressed
within walls is found to be similar to the mean velocity of walls
and of all particles, $\langle w_w\rangle\sim\sigma_{v}\approx
\sigma_{vel}$. In contrast, for the SCDM and $\Lambda$CDM models
the dispersion $\langle w_w\rangle$ is  about 30\% smaller
than that obtained for the complete walls in DMRT and the
dispersions  $\sigma_{vel}$ and $\sigma_v$ discussed above. This
divergence characterizes statistically the evaporation of high
energy particles in course of the relaxation of compressed matter
and is reinforced by the procedures of measurement and wall
selection. The relatively small value of $\langle w_w\rangle$
demonstrates that in contrast to the clusters of galaxies the
moderate degree of one dimensional matter compression within walls
is not accompanied by an essential growth of velocity dispersion.

\subsubsection{Relaxation of compressed matter}

For $\Lambda$CDM and SCDM models the wall thickness, $h_w\sim$
(3 -- 4)$h^{-1}$Mpc, is 2 -- 3 times smaller than that expected
in the Zel'dovich theory (3.10). So large compression of matter
within walls means that the selected particles are strongly confined
and, probably, relaxed. For 1D matter compression the relaxation is
expected to be weak, but in reality it is reinforced due to the small
scale clustering and disruption of walls.

\begin{figure}
\centering
\epsfxsize=7.5 cm
\epsfbox{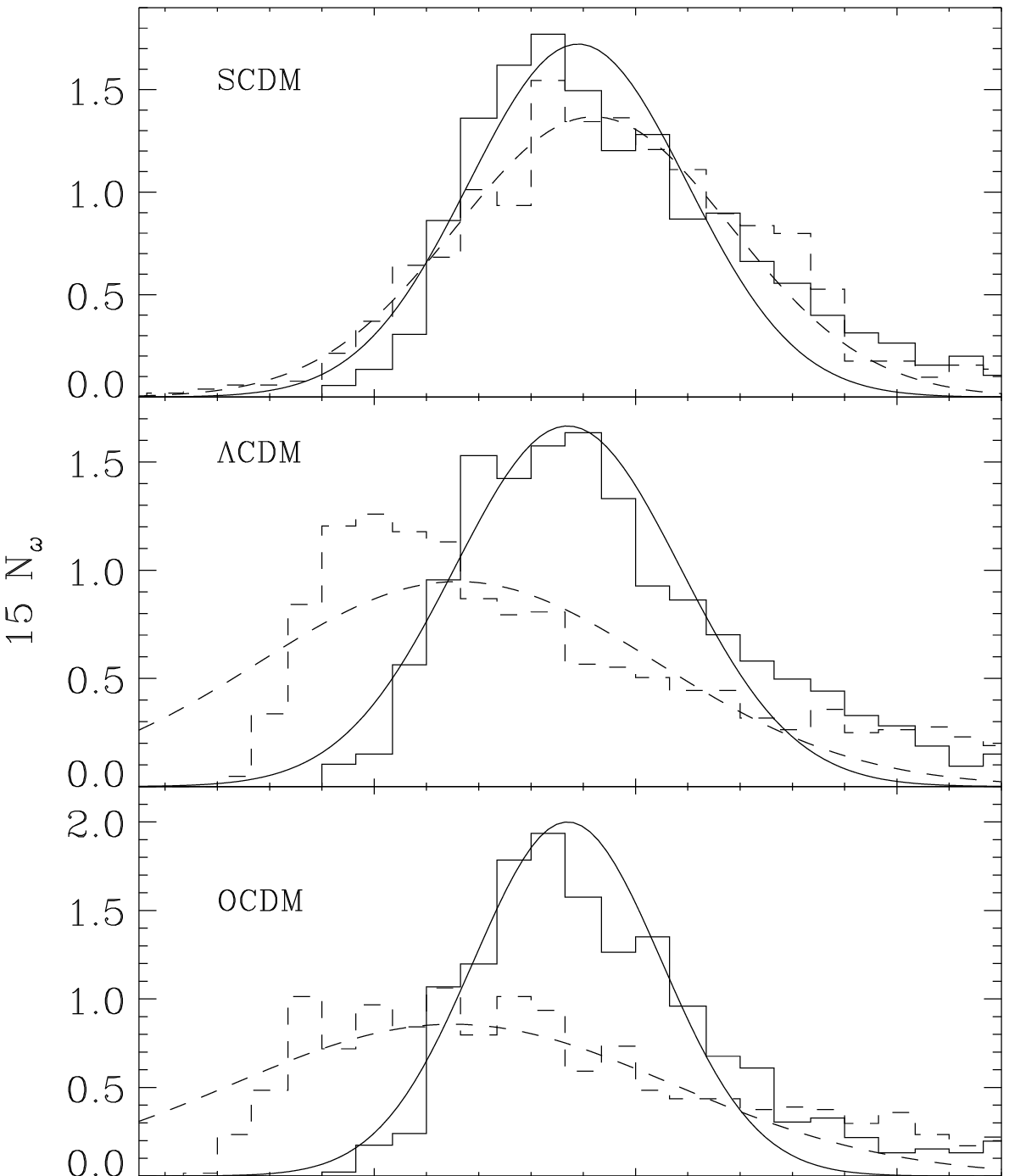}
\vspace{0.75cm}
\caption{PDFs of the reduced velocity dispersion,
$N_\omega(\omega/\langle\omega\rangle)$, for DM walls in real
(solid lines) and redshift (dashed lines) spaces for SCDM,
$\Lambda$CDM and OCDM models. The Gaussian fits are also shown
by solid and dashed lines.
}
\end{figure}

\begin{table*}
\begin{minipage}{180mm}
\caption{Wall characteristics in real and redshift spaces.}
\label{tbl2}
\begin{tabular}{ccc ccc ccc ccc cc } 
sample&$\delta_{thr}$&$f_w$&$f_{cr}$&$\langle q_w\rangle$&
$\langle\tau_m\rangle$&$\langle\tau_v\rangle$&$\langle\delta\rangle$&
$\langle h_w\rangle$&$\langle w_w\rangle$&$\langle\epsilon\rangle$&
$p_w$&$\langle D_{sep}\rangle$&$f_{dq}$\cr
& & & & & & & &$h^{-1}$Mpc&km/s & & &$h^{-1}$Mpc&\cr
\multicolumn{14}{c}{real space}\cr
        SCDM &2.5&0.44&0.74&$1.00\pm 0.18$&$0.39\pm 0.04$&$0.58$&4.7&$~~3.4\pm
0.4$&$463\pm 18$&0.1&0.27&$43\pm ~~8$&0.4\cr 
$\Lambda$CDM &1.6&0.46&0.85&$0.83\pm 0.16$&$0.35\pm 0.03$&$0.39$&7.4&$~~4.0\pm
0.3$&$387\pm 20$&0.2&0.30&$71\pm  13$&0.4\cr 
        OCDM &1.3&0.40&0.88&$0.52\pm 0.06$&$0.28\pm 0.02$&$0.26$&2.4&$~~6.0\pm
0.7$&$330\pm 33$&0.8&0.44&$46\pm ~~8$&0.3\cr 
mock$_1$     &1.6&0.43&0.82&$0.87\pm 0.15$&$0.36\pm 0.03$&$0.27$&5.0&$~~7.0\pm
1.1$&$412\pm 51$&0.9&0.48&$84\pm 17$&0.3\cr  
mock$_2$     &1.3&0.44&0.81&$0.84\pm 0.13$&$0.36\pm 0.03$&$0.27$&5.1&$~~6.6\pm
0.9$&$394\pm 45$&0.8&0.47&$78\pm 13$&0.3\cr  
mock$_3$     &1.3&0.45&0.84&$0.88\pm 0.11$&$0.36\pm 0.02$&$0.27$&5.6&$~~5.2\pm
0.6$&$354\pm 27$&0.6&0.43&$73\pm 11$&0.4\cr  
mock$_4$     &1.3&0.44&0.86&$1.23\pm 0.17$&$0.43\pm 0.03$&$0.28$&9.1&$~~4.5\pm
0.4$&$359\pm 20$&0.5&0.39&$78\pm 14$&0.5\cr  
mock$_4$     &1.3&0.48&0.87&$1.35\pm 0.19$&$0.45\pm 0.03$&$0.27$&7.5&$~~5.6\pm
0.5$&$379\pm 26$&0.7&0.40&$81\pm 12$&0.5\cr 
\hline
\multicolumn{14}{c}{redshift space}\cr
        SCDM &2.5&0.41&0.88&$0.84\pm 0.08$&$0.36\pm 0.02$&  --  &1.8&$~~8.5\pm
0.8$&$245\pm 23$&--&0.75&$45\pm ~~9$&0.3\cr 
$\Lambda$CDM &2.1&0.45&0.88&$0.76\pm 0.13$&$0.34\pm 0.03$&  --  &2.6&$~~7.2\pm
0.6$&$207\pm 17$&--&0.48&$63\pm  11$&0.4\cr 
        OCDM &1.3&0.44&0.77&$0.56\pm 0.10$&$0.29\pm 0.03$&  --  &1.4&$ 11.2\pm
1.3$&$323\pm 36$&--&0.58&$49\pm ~~8$&0.3\cr 
mock$_1$     &1.2&0.43&0.82&$0.89\pm 0.16$&$0.37\pm 0.03$&   -- &2.7&$ 13.3\pm
2.3$&$385\pm 66$&--&0.71&$87\pm  18$&0.3\cr  
mock$_2$     &1.5&0.43&0.83&$0.86\pm 0.13$&$0.36\pm 0.03$&   -- &2.7&$ 12.2\pm
1.9$&$353\pm 54$&--&0.70&$86\pm  25$&0.3\cr  
mock$_3$     &1.8&0.44&0.86&$0.85\pm 0.09$&$0.36\pm 0.02$&   -- &2.9&$~~9.5\pm
1.1$&$276\pm 32$&--&0.70&$83\pm  17$&0.3\cr  
mock$_4$     &1.8&0.45&0.84&$1.10\pm 0.13$&$0.41\pm 0.02$&   -- &4.1&$~~8.8\pm
0.9$&$254\pm 25$&--&0.65&$87\pm  21$&0.4\cr  
mock$_4$     &1.3&0.46&0.82&$1.22\pm 0.13$&$0.43\pm 0.02$&   -- &3.8&$ 10.1\pm
0.9$&$291\pm 26$&--&0.61&$89\pm  16$&0.4\cr  
\hline
\end{tabular}

Here $\delta_{thr}$ is threshold parameters of clusters, $f_w$ is 
the fractions of all particles forming the selected walls, and 
$f_{cr}$ is the fraction of particles belonging to walls in both 
real and redshift spaces. Parameters $\tau_m, \tau_v$ are the 
amplitude of perturbations as given by (3.1)~\&~ (3.4). The other 
quantities are explained in the text. Averaging was performed 
over 7 core sizes, $6h^{-1}$Mpc$\leq D_{core}\leq 9h^{-1}$Mpc, and
over 7 threshold richness, 10$\leq N_{min}\leq$35 for the
linking length $l_{link}=5h^{-1}$Mpc.
\end{minipage}
\end{table*}

The degree of relaxation reached can be characterized by
the parameters
$\langle \delta\rangle  ~\&~\langle \epsilon\rangle$,
$$\langle \delta\rangle  = \left\langle {l_v q_w\over h_w}\right
\rangle,\quad \langle\epsilon\rangle ={\langle w^2_w\rangle\over
w_z^2(\tau)},				\eqno(6.3)$$
listed in Table 2. Here $\langle \delta\rangle $ measures the
mean degree of matter compression, and $\langle\epsilon\rangle$
is the mean kinetic energy of compressed particles with respect
to the expectations of the Zel'dovich theory. The function
$w_z(\tau)$ given by (3.7) is evaluated at $\tau=\tau_v$.

The divergence between the expectations of Zel'dovich theory
and simulations is moderate for the OCDM model and becomes strong
for the $\Lambda$CDM model as the evolution progresses. For the SCDM
model the estimate of $\langle\delta\rangle$ is artificially
decreased together with $\langle q_w\rangle$. The small value
of $\langle\epsilon\rangle\sim$ 0.1 -- 0.2 confirms an essential
deficit of energy of compressed particles in comparison to that
expected in the Zel'dovich theory. This deficit is partly
enhanced by the procedure of wall selection, as the wall
boundaries are blured, and particles placed far from the wall
center are not included into walls.

In Zel'dovich theory the strong correlation of $w_w$ and $h_w$
with the wall richness, $m_w$, is described by expressions (3.6)
\& (3.9). In simulations the measured linear correlation
coefficients of $q_w$,  $w_w$ and $h_w$ are also $\sim$ 0.4 --
0.5 what indicates that the essential mass dependence of these
parameters  remains also after relaxation. To discriminate the regular
and random variations of functions $w_w$ and $h_w$ we will
consider the {\it reduced} wall thickness, $\zeta$, and the
{\it reduced} velocity dispersion, $\omega$, which can be defined
as follows:
$$h_w = \langle h_w\rangle\mu^{p_h}\zeta,\quad
w_w = \langle w_w\rangle\mu^{p_w}\omega,       \eqno(6.4)$$
$$\mu = m_w/\langle m_w\rangle = q_w/\langle q_w\rangle,\quad
p_h\approx p_w\approx 0.3 - 0.4,$$
$$\langle\zeta\rangle\approx\langle\omega\rangle\approx 1,\quad
\sigma_\zeta\approx\sigma_\omega\approx 0.2.$$

In all considered cases the PDFs of the reduced velocity dispersion
within walls, $N_\omega$, and of the reduced wall thickness,
$N_\zeta$, can be roughly fitted to Gaussian functions. The PDFs
$N_\omega$ are plotted in Fig. 4 for all three models.

These results show that due to the strong relaxation of
compressed matter the correlations between the considered
characteristics of walls predicted by the Zel'dovich theory
in Eqs. (3.6), (3.8) \& (3.10) are replaced by relations (6.4)
which are also universal.

\subsection{DM walls in the redshift space}

If the analysis of wall characteristics in the real space allows
to reveal the influence of gravitational interaction of the compressed
matter, then a similar analysis performed in the redshift space
reveals the influence of random velocities on the observed
characteristics of the large scale matter distribution.

In the redshift space the analysis of wall characteristics was
performed for samples of walls selected as described in
Sec. 6.1 . As was shown in DMRT, in low density cosmological
models the main characteristics of these walls are similar
to the observed characteristics of superclusters of galaxies.
The determination of wall characteristics and their corrections
are discussed in Sec. 4.3. The wall parameters were found in the
same ranges of $D_{core}$ and $N_{min}$ as in the real space for
$l_{link}=5h^{-1}$Mpc.

The main results are listed in Table 2 and are plotted in Figs.
2 -- 4.

\subsubsection{Walls in the real and redshift spaces}

The samples of walls selected in the real and redshift spaces are not
identical with each other due to influence of random velocities of
particles. This difference can be suitably characterized by the
fraction of the same particles assigned to walls in both spaces.
Here this fraction was defined as a ratio of number of the 
particles, $N_{com}$, to the number of particles assigned to
the selected walls, $N_w$. For all models under consideration this
fraction, listed in Table 2, is
$$f_{cr} = N_{com}/N_w\sim0.8 - 0.9.$$
Small variations of number of particles, $N_w$, assigned to walls
in the real and redshift spaces lead to these variations. 

These results indicate that the influence of high random velocities
generated by the small scale wall disruption and the matter relaxation
moderately distorts the sample of walls selected in the redshift space.
More strong deviations between such wall parameters as the wall
thickness and degree of matter compression, measured in the real and
redshift spaces, are caused by the redistribution of matter within
walls and the procedure of measurement rather than by the incorrect
wall identification. The impact of these factors rapidly increases with
$\tau_m$ and becomes extreme for the SCDM model.

These deviations can be sensitive to the code used for simulation
(see, e.g., discussion in Splinter et al. 1998). For example,
in the P$^3$M code, these variations may increase due to
larger velocities of compressed matter generated there.

\subsubsection{Basic characteristics of DM walls}

For all three models the mean surface density of selected walls
listed in Table 2 is similar to that found in the real space. This
fact shows that the artificial growth of matter concentration within
walls discussed in Sec. 3.2  is effectively suppressed by the
influence of the velocity dispersion and the procedure of wall
selection, and the relation (3.1), as before, connects the mean
surface density of selected walls, $\langle q_w\rangle$, with the
amplitude, $\tau_m$. Variations of $\langle q_w\rangle$ and $\tau_m$
with $D_{core}$ and $N_{min}$ are shown in Table 2 as a scatter of
these parameters. The PDFs $N_m$ plotted in Fig. 2 are also similar
to those found in the real space.

The mean wall separation is consistent with the estimate found in
real space and, as before, for all models $f_{dq}\approx
(0.75 - 0.9)f_w$. The PDFs $N_{sep}$ plotted in Fig. 3
are also similar to those found in the real space.

In the redshift space the used method of wall identification selects
mainly particles with a small relative velocity what essentially
restricts the measured velocity dispersion within walls and the
wall thickness. Results listed in Table 2 show that only for the
OCDM model, the velocity dispersion of compressed matter is
consistent with the values found in the real space and in DMRT.
For $\Lambda$CDM and SCDM models they are even smaller than those
found in the real space. The measured wall thickness is now linked with
the velocity dispersion by the relation (3.16).

\subsubsection{Characteristics of matter relaxation}

In the redshift space walls are less conspicuous than in the real
space but, even so, for all three models the mean overdensity,
$\langle\delta\rangle$, listed in Table 2, differs from the
estimates based on the  Zel'dovich theory (3.7). As in the real space, the
velocity dispersion in the redshift space is strongly correlated with
the surface density of walls, what is described by the relation (6.4)
with an exponent $p_{w} \approx$ 0.5. The PDFs of the reduced
velocity dispersions, $N_\omega$, plotted in Fig. 4, demonstrate
some excess of particles with lower $\omega$, but it can also be
roughly fitted to a Gaussian function with $\langle \omega\rangle
\approx$1 and dispersion $\sigma_\omega\approx 0.4$. This dispersion
is about two times larger than that in the real space.

These results show that, although in the redshift space walls are
not so conspicuous as in the real space, in the range of `time'
$\sim 0.2\leq\tau\leq~\sim 0.5$, the relaxation of compressed
dark matter can be directly recognized with these methods.

\subsection{Walls in mock catalogues}

The analysis of mock catalogues characterizes how the considered
simple model of large scale bias influences the measured wall
properties. These catalogues were investigated also both in
the real and redshift spaces. The analysis was performed for 10 
values $N_{min}$ (15$\leq N_{min}\leq$60) and for 7 values of the 
core radius $D_{core}$ (7$h^{-1}$Mpc$\leq D_{core}\leq 10h^{-1}$Mpc) 
using a linking length
$l_{link}=5h^{-1}$Mpc. The main results averaged over these
$N_{min}$ and $D_{core}$ are listed in Table 2.

\begin{figure}
\centering
\vspace{0.05cm}
\epsfxsize=7 cm
\epsfbox{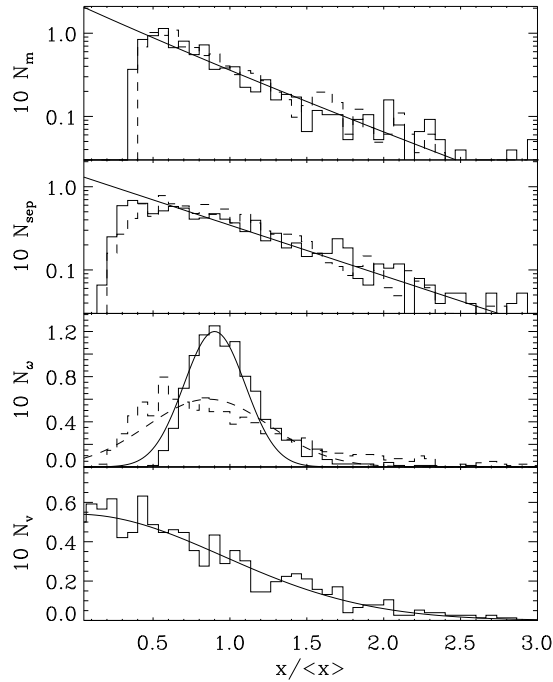}
\vspace{0.8cm}
\caption{PDFs of wall surface density, $N_m(q_w/\langle
q_w\rangle)$, wall separation, $N_{sep}(D_{sep}/\langle D_{sep}
\rangle)$, reduced velocity dispersion, $N_\omega(\omega/\langle
\omega\rangle)$, and velocity of walls, $N_v(v_w/\sigma_{v})$,
for the $mock_4$ catalogues in the real (solid lines) and
redshift (dashed lines) space. The same fits as in Figs. 1 -- 4
are plotted as well.
}
\end{figure}

\subsubsection{'Galaxy' walls in the real space}

In the real space for all mock catalogues the parameters $\tau_v$,
$\langle h_w\rangle$ and $\langle \epsilon_w\rangle$ are similar
to those found for the basic OCDM model. The velocity dispersion
of `galaxies' within walls, $\langle w_w\rangle\approx\sigma_v
\approx\sigma_{vel}$, exceeds that found for the basic model by
about of 20 -- 30\%. These variations can be attributed to the
preferential identification of `galaxies' in the central high
density regions of walls, where the relative velocities of DM
particles are also larger than the mean values. The wall thickness
and the velocity dispersion of `galaxies' can be reduced and
turned into dimensionless quantities in the same manner as in
Eq. (6.4), and the PDFs for the reduced wall thickness and velocity
dispersion within walls, $N_\zeta$ and $N_\omega$, are also
similar to Gaussian functions. The PDFs $N_\omega$ are shown in
Fig. 5 for the mock$_4$ catalogue.

 As was expected, the mean surface density of walls, $\langle
q_w\rangle$, exceeds that found for the basic  OCDM model, and
this excess progressively increases together with the biasing
factor used. This excess can be considered as a suitable measure
of the bias. This means that to characterize this bias the
difference between $\tau_m$ and $\tau_v$ and/or between $\tau_m$
and other amplitudes measured for the same catalogues can be
used together with the autocorrelation function. The growth
of $q_w$ leads to a proportional growth of $\delta$, as the
wall thickness is only weakly distorted.

The large scale bias increases the contrast between richer and
poorer walls what is seen as an essential growth of the mean wall
separation. In all mock catalogues $\langle D_{sep}\rangle$ is
about two times larger than in the basic OCDM model. The
growth of both $\langle q_w\rangle$ and $\langle D_{sep}\rangle$
do not distort the relation between $f_w$ and $f_{dq}$.

\subsubsection{'Galaxy' walls in the redshift space}

In the redshift space the fraction of the same particles assigned
to walls both in the real and redshift spaces becomes $f_{cr}\sim$ 80
-- 85\% (Table 2), what explains the similarity of parameters
$\langle q_w\rangle$ and $\tau_{m}$ listed in Tables 2 for both
cases. The expected growth of wall richness in the redshift space
according to (3.13) is not found, and the surface densities of
walls, $\langle q_w\rangle$, are, in the range of errors, the
same both in the real and redshift spaces. This fact shows that for
`galaxies' the expected growth of wall richness in the redshift
space is suppressed even more strongly than for DM component due
to the relaxation of compressed matter. Then Eq. (3.1) describes
correctly the time dependence of the mean wall surface density.

The parameters $\langle h_w\rangle ~\&~\langle w_w\rangle$
for `galaxy' walls are similar to those found for the
underlying OCDM model. The difference of $\langle w_w\rangle$
found for the same samples of walls in the real and
redshift spaces and a slow decrease of $\langle w_w\rangle$
for stronger biased models can be assigned to the loss of
a small fraction of particles with large velocities, what
demonstrates the sensitivity of these functions to the method
of wall identification.

In the redshift space we have not so a reliable independent estimator
of the amplitude as $\tau_v$. There the bias is seen as a relation 
of the amplitudes $\tau_m\geq\tau_\xi$. This makes it difficult to
quantitatively estimate  the relatively moderate large scale bias
in observed catalogues because both $\tau_m~\&~\tau_\xi$ are
sensitive to the bias.

This discussion shows that the simple algorithm used in DMRT
for the `galaxy' identification does not essentially distort the
basic characteristics of simulated walls, and a stronger
bias can be seen as an excess of the surface density of `galaxies'
relative to that found for the DM component in the basic model.
At the same time the mean velocity dispersions of both the DM component
and the `galaxies' assigned to walls, $\langle w_w\rangle$, tends to
be smaller than $\sigma_v~\&~\tau_v$ and other characteristics of
the amplitude of perturbations.

\section{Summary and discussion}

In this paper we continue the investigation of large scale
matter distribution and processes of large scale structure
formation and evolution. Some aspects of these problems were
discussed in our previous papers (LCRS-1, LCRS-2, DURS, DMRT,
M\"uller et al. 1998) where the 3-dimensional analysis of
the observed and simulated large scale structure was performed
with the core-sampling and the Minimal Spanning Tree techniques.
Another approach to this problem, based on the percolation
technique, was discussed in Sahni et al. (1994), Shandarin
\& Yess (1998) and Sathyaprakash et al. (1998). The
statistical description of structure formation and evolution
based on the Zel'dovich theory of nonlinear gravitational
instability can be found in Lee \& Shandarin (1998) and DD99.

Here we direct our attention to the physical aspects
of the process of wall formation, what implies a more detailed
discussion of the properties of DM walls in real space. The
simulations described and investigated in DMRT are used to test
the theoretical expectations, to estimate the influence of
small scale clustering and relaxation of compressed matter
and other random factors, and to examine the power of
the statistical methods used to describe the large
scale matter distribution. Three cosmological models, at
different evolutionary stages, were analyzed
in the same manner, and the comparison of results obtained
for these models allows us to estimate the properties of
walls at various $\tau$.

In the redshift space the influence of small scale clustering and
large velocity dispersion of compressed matter noticeably
distorts some characteristics of the walls. These distortions
appear also in the considered mock catalogues, and can even be
enhanced by the possible large scale bias between the spatial
distribution of DM and galaxies.

Some of these results may depend on the code used for the
simulations (see, e.g., the discussion in Splinter et al. 1998),
and they should be checked with simulations employing
a code with higher spatial resolution.

\subsection{Identification of walls}

The core-sampling approach described in Sec. 5 allows us to
characterize, in more details, the matter distribution along
the sampling core and to estimate the uncertainty in measured
properties of wall-like condensations introduced by the
influence of velocity dispersion and small scale clustering.
The influence of these random factors is demonstrated by
comparing results obtained with various $D_{core}, N_{min}~\&~
l_{link}$.

Results presented in Sec. 6 show that some fraction of
the early compressed matter has subsequently evaporated due
to relaxation processes. These DM particles together
with the infalling matter form an extended halo around the walls and,
therefore, it is difficult to separate the walls from the
background. The same problem  is met by the correct
definition of boundaries of galaxies and  clusters
of galaxies. It was also discussed in the DMRT, LCRS-2 and
DURS, where the methods of wall selection, described in Sec.
6, were applied to simulated DM and observed galaxy
distributions.

The central high density part of walls is reliably selected in all the
cases, but various definitions of the wall boundaries can noticeably
change the measured characteristics of walls. To provide more
objective comparisons of wall characteristics the same dimensionless
parameters $f_w$ and $\delta_{thr}$ should be used for
identification of walls in different catalogues and simulations.

\subsection{DM walls in the real space}

\subsubsection{Measured characteristics of walls}

The results presented in Sec. 6 show that the core-sampling approach
can be successfully used for the investigation and description
of the large scale matter distribution and the wall-like matter
condensations. It allows to estimate the surface density,
thickness, velocity dispersion and other basic parameters of
DM walls corrected for the influence of random curvature and
shape of walls. These parameters differ from those obtained
in 3D space with the Minimal Spanning Tree and inertia tensor
methods, and these methods suitably complement each other.

The measured wall characteristics can be compared with
predictions of the Zel'dovich theory what reveals the
influence of relaxation of compressed matter on the properties
of walls and allows to correct the theoretical expectations.
The small scale clustering of compressed matter and the wall
disruption lead to noticeable variations of measured wall
characteristics for different parameters of the sampling
core. These variations are not so large for the low
density models, but they increase rapidly with $\tau$.

The dimensionless surface density of walls, $q_w$, is closely
connected with the size of proto-walls as discussed in DMRT,
LCRS2 and DURS. The high surface density of
walls, $q_w\geq 0.6$, found above even for the low density
models, demonstrates that processes of strong nonlinear
matter evolution occur at a typical scale of $\sim
q_wl_v\sim$ (15 -- 25)$h^{-1}$Mpc. This evolution is
correctly described by the Zel'dovich theory. This
characteristic is sensitive to the basic cosmological
parameters, $\Omega_m ~\&~h$, what allows us to select the
class of more perspective models for further investigation.

\subsubsection{Relaxation of compressed matter}

The problem of relaxation of compressed matter is now
in the forefront, and the obtained results allow to begin
discussion of the statistical characteristics of this
relaxation. The analysis performed in the real space is more
important for the discussion of the basic physical processes
which occurred during the formation of wall-like matter
condensations, such as the small scale matter clustering and
the relaxation of the compressed matter. These processes
generate the large velocity dispersion within walls and lead
to the evaporation of high velocity particles. Thus, in all
these cases an essential deficit of energy in DM walls as
compared with the expectations of the Zel'dovich theory --
$\sim$ (50 -- 80)\% and more -- was found. The growth of this
deficit with $\tau$ from the OCDM to SCDM models demonstrates
that the DM relaxation becomes more and more important for
later evolutionary stages, and its influence on the observed
parameters of the large scale matter distribution becomes
crucial for $\tau \geq 0.5$.

The relaxation is seen in rich superclusters of galaxies
such as the Perseus-Pisces (Saslaw \& Haque-- Copilah 1998).
It is essentially accelerated and amplified by
the small scale clustering of compressed matter. This
clustering is clearly seen in observations as, for example,
a strongly inhomogeneous galaxy distribution within the
Great Wall (Ramella et al. 1992). The clusters of galaxies
situated within wall-like superclusters similar to the Great
Wall and the Perseus-Pisces can be considered as extreme
examples of this process.

The merging of earlier formed structure elements is very
important for the formation of large walls (DD99). This means
that actually the relaxation occurs step by step during all
the evolutionary history beginning with the formation of first
low mass, high density pancakes which later are successively
integrated and merged to larger structure elements. This means
also that the finally reached degree of relaxation and the
properties of compressed matter depend on the (unknown)
evolutionary history of the considered walls and, therefore,
can be characterized only statistically.

The relaxation of compressed matter destroys the tight
correlation between the surface density and velocity dispersion
predicted by the Zel'dovich theory (3.6), but it generates other
correlations between the same characteristics described by
the relations (6.4). This fact indicates that the
properties of compressed matter are sufficiently general, and
these characteristics can be used to improve the methods
of wall selection and the description of wall properties.

The velocity dispersion within walls increases gradually
with $\tau$ from the OCDM to the SCDM model. As was discussed
in Sec. 7, the particles with high velocity are gravitationally
confined and occupy preferentially the high density central
regions of walls. This fact confirms that, probably, these
particles are relaxed and have a (quasi)stationary distribution.
This distribution is not so stationary as, for example, in
clusters of galaxies, and it is slowly evolving due to the
large scale matter flow along the walls and the persisting
merging of neighboring structure elements, but presumably, this
evolution does not significantly distort the formed matter
distribution.

\subsection{DM walls in the redshift space}

The matter condensation seen in the redshift space can be partly
artificially enhanced by the influence of streaming velocities.
The possible influence of this effect was widely discussed
over the past decade (see, e.g., Kaiser 1987, McGill 1990 a;
Davis, Miller \& White 1997; Hamilton 1998; Hui, Kofman \&
Shandarin 1999) and, as applied to properties of absorption
lines in the spectra of high redshift quasars, by McGill (1990b)
and more recently by Levshakov \& Kegel (1996, 1997). These
tendencies are also clearly seen from the direct application
of the Zel'dovich approximation to the wall formation in the
redshift space as was discussed in Sec. 3.2. Of course, it is
impossible to decide which particles belong to walls, but we
can estimate statistically the properties of DM walls identified 
in the redshift space. However, the influence of this uncertainty 
cannot be separated from the influence of the relaxation and other
factors discussed above.

For all models the comparison of DM walls selected in the real
and redshift spaces demonstrates, that they are composed
mainly from the same particles -- this fraction is about
$f_{cr}\sim$ (70 -- 80)\% (Table 2). This means that in both
cases we find the same walls, and the fraction
of randomly added or lost particles is indeed small.
In spite of this, some properties of walls in the redshift space
are quite sensitive to the velocity dispersion and to the
methods of wall identification. Thus, the strong growth of
wall thickness -- about a factor of 2 -- confirms results
obtained by Melott et al. (1998). This effect is quite
similar to the well known 'finger of God' effect observed
in clusters of galaxies.

The wall surface density, $q_w$, is most interesting, as
it is directly connected with the basic cosmological parameters,
$\Omega_m~\&~ h$. Our analysis shows that for low density
models -- $\Lambda$CDM and OCDM -- the measured value of $q_w$
is similar both in the real and redshift spaces. This means
that the growth of the matter condensation within walls
due to streaming velocities as predicted by the Zel'dovich theory
is strongly suppressed by the influence of the matter relaxation 
and the transformation of a continuous matter infall to a 
discontinuous one. Actually similar relations connect the 
fundamental wall characteristics such as $q_w$ and $\tau_m$.

The velocity dispersion within walls selected in the 3D  redshift
space can be noticeably underestimated what is a direct
consequence of the method of wall selection. In the redshift space,
particles with large velocities are artificially shifted to
the periphery of selected walls and, so, can be omitted from the
analysis.

\subsection{Walls in mock catalogues}

For the considered mock catalogues the influence of velocity
dispersion is enhanced by the methods used for `galaxy'
selection. The large scale bias increases the `galaxy'
concentration within walls and, so, increases the density
gradient near the wall boundary. When the `galaxies' are
identified preferentially in the high density central parts
of the walls (in the real space), than their velocity
dispersion exceeds that for the DM particles, and this excess
may be as large as $\sim$ (20 -- 30)\%. In the redshift space,
the parameters of `galaxy' walls such as $\langle h_w\rangle$ and
$\langle w_w\rangle$ are similar to those in the underlying
DM distribution.

The bias is clearly seen both in the real and redshift
spaces as an excess of the mean surface density of walls. The
comparison of parameters $q_w~\&~\tau_m$ found for observed
wall-like galaxy condensations with possible independent
estimates of the same parameters gives us a chance to obtain
a reasonable observational estimates of the large scale bias.

\subsection{The amplitude of large scale perturbations}

These results demonstrate again that all characteristics of
the amplitude and evolutionary stage of large scale structure
considered in Secs. 2 \& 3 are similar, but not identical to
each other, as they are sensitive to different properties of
perturbations. The best and most stable measure, $\tau_v$,
comes from measurements of the velocity of structure elements.
It is insensitive to the nonlinear evolution of perturbations,
large scale bias and small scale clustering or relaxation of
the compressed matter.

The comparison of other estimates for the same parameter
$\tau$, namely, $\tau_{vel}, ~\tau_\xi,~\&~\tau_m$ obtained
in the same simulations demonstrates their sensitivity
to various natural and artificial factors. For the low density
models -- $\Lambda$CDM and OCDM -- the parameters $\tau_v$ and
$\tau_m$ are usually sufficiently close to each other, what
is a direct consequence of the close connection of the process
of wall formation with the large scale perturbations. The
parameter $\tau_m$ is sensitive to a possible large scale bias,
but to reveal this factor, we need to have independent unbiased
estimates of the same amplitude.

The most interesting independent estimate of the amplitude
is $\tau_\xi$ which is however more sensitive to small scale
matter clustering. Thus, for the SCDM model where this clustering
is stronger it significantly overestimates the large scale
amplitude. It is less sensitive to the large scale
bias  than $\tau_m$.

Independent estimates of the large scale amplitude come
from measurements of the CMB anisotropy. The COBE data are
consistent with other available estimates of cosmological
parameters and of the large scale amplitude, and therefore,
$\tau_T$ can be considered as the best estimate of the
combination (2.9) of $\Gamma$ and the amplitude. It can
be connected with estimates of cosmological parameters
$\Omega_m\approx 0.3,~~\Omega_\Lambda\approx 0.7$ obtained
from observations of high-redshift supernovae (Perlmutter et
al. 1998). Nonetheless, $\tau_T$ should be corrected for a
possible contribution of gravitational waves.

The investigation of the space density of clusters of galaxies
and its redshift evolution (see, e.g., Bahcall \& Fan 1998;
Eke et al. 1998; Wang \& Steinhardt 1998) seems also to be
promising and can give the required independent measure
of the large scale amplitude. The formation and evolution of
galaxy clusters is caused by large scale perturbations, and
their characteristics can be connected with these
perturbations. But they are sensitive to the thermal
evolution of clusters and, moreover, are related to only
$\sim$ (10 -- 15)\% of matter accumulated by the clusters.
This means that they are not free from random variations what
is seen, in particular, as the well known variations of the
autocorrelation function with the cluster sample.

The critical discussion of available measurements of
cosmological parameters (Wang et al. 1999; Efstathiou 1999)
shows that in spite of a large progress reached during last
years, we do not have yet a reliable unbiased estimate of
these parameters, and these data should be tested with respect
to possible random large scale variations. The application
of the discussed methods to large observed redshift surveys
can help to achieve this goal.

\subsection*{Acknowledgments}
We are grateful to our anonymous referee for the useful comments and 
criticism. 
This paper was supported in part by Denmark's Grundforskningsfond
through its support for an establishment of Theoretical Astrophysics
Center and Polish State Committee for Scientific Research grant Nr.
2-P03D-014-17. AGD also wishes to acknowledge support from the Center
for Cosmo-Particle Physics "Cosmion" in the framework of the project
"Cosmoparticle Physics".

\bigskip\bigskip\bigskip

\centerline{\bf Appendix A}
\medskip
\centerline{\bf Dynamical characteristics of walls}
\centerline{\bf in the Zel'dovich theory}
\medskip

\noindent
The results obtained in DD99 allow us to discuss in more details
dynamical characteristics of walls predicted in the Zel'dovich theory.
The comparison of these expected and actually simulated characteristics
reveals the influence of interaction and relaxation of compressed matter.

Following  DD99 we define the wall formation as the intersection
of two DM particles with different Lagrangian coordinates, ${\bf q}_1$
and ${\bf q}_2$. The difference of these coordinates measures the
size of the pancake. Using the basic relations of the Zel'dovich theory
(2.2) and (2.3), linking the Lagrangian and Eulerian coordinates
and velocities of particles, we obtain the coordinate and velocity
of a wall as a whole (DD99):
$$r_w={1\over l_vq_w}\int_{q_1}^{q_2}{\bf nr}~dq =
{l_v\over 1+z}\left(q_c-\tau(z){\Delta \Phi\over q_w}\right),\eqno(A.1)$$
$$v_w = {1\over l_vq_w}\int_{q_1}^{q_2} {\bf n}[{\bf u}
-H{\bf r}]dq = $$
$${l_vH(z)\over 1+z}\left[q_c-\tau(z)(1+\beta)
{\Delta\Phi(q_w)\over q_w}\right],$$
$${\bf n}={ {\bf q}_1 - {\bf q}_2\over |{\bf q}_1 - {\bf q}_2|},\quad
q_c={|{\bf q}_1 + {\bf q}_2|\over 2l_v},\quad
q_w={|{\bf q}_1 - {\bf q}_2|\over l_v},$$
where
$\Delta\Phi(q_w)$ is the random difference of the dimensionless
gravitational potential over the wall. It is convenient to
introduce the relative normalized Lagrangian coordinate of a
particle within a wall, $\vartheta$:
$$q_p=q_c+0.5 q_w\vartheta,\quad -1\leq \vartheta\leq 1.$$
Using the coordinate $\vartheta$ we will describe the relative
position and velocity of the infalling particle with the Lagrangian
coordinate $q_p$ or $\vartheta$ by the functions:
$$r_{inf}={\bf nr}-r_w = {l_v\over 1+z}\left[{q_w\over 2}\vartheta-
\tau(z)\left(~S(\vartheta) - {\Delta\Phi(q_w)\over q_w}\right)\right],$$
$$v_{inf}={\bf nv}-v_w = -u(z) 0.5 q_w\vartheta +H(z)(1+\beta)r_{inf},
						\eqno(A.2)$$
$$u(z) = H(z)l_v\beta(z)(1+z)^{-1}.$$
Here $S={\bf nS}$ is thee random dimensionless longitudinal displacement
of a particle from its unperturbed Lagrangian position introduced by
(2.2).

For Gaussian initial perturbations the PDF of the random function
$r_{inf}$ is also Gaussian, and the mean value and dispersion of
$r_{inf}$ should be found using the conditional characteristics
of functions $S$ and $\Delta\Phi$ taking into account that a wall
is formed in the point $r=r_w$ (DD99). In this case for walls with
$q_w\ll$1 we have:
$$\langle r_{inf}\rangle\approx {l_v\over 1+z}{q_w^3\over 4}\vartheta
\ll\sqrt{\langle r_{inf}^2\rangle}\approx {l_v\tau(z)\over 1+z}
\sqrt{q_w\over 3},			\eqno(A.3)$$
and $\langle r_{inf}^2\rangle$ is independent from $\vartheta$.
This means that both random functions,
$$r_{inf}~\&~v_{inf}+u(z)0.5q_w\vartheta = H(z)(1+\beta)r_{inf}$$
are also independent from $\vartheta$. Hence, for the thickness,
$h_w$, of a wall with the surface density $q_w$, and for the velocity
dispersion within such a wall we have
$$h_w^2 = 12\cdot{1\over 2} \int_{-1}^1 d\vartheta~\langle
r_{inf}^2\rangle =4l_v^2\tau^2q_w(1+z)^{-2},		\eqno(A.4)$$
$$w_w^2 = {1\over 2}\int_{-1}^1 d\vartheta~\langle v_{inf}^2
\rangle={H^2l_v^2\over (1+z)^2}\left({\beta^2\over 12}q_w^2+
{\tau^2(1+\beta)^2\over 3}q_w\right).$$
Here the wall thickness is normalized by the thickness of a 
homogeneous slice.

\end{document}